\documentclass[12pt]{article}

\usepackage{graphicx}
\usepackage{times}
\usepackage{scicite}

\usepackage{multirow}

\newenvironment{sciabstract}{%
\begin{quote} \bf}
{\end{quote}}

\usepackage{color,soul}
\usepackage{amsmath}
\usepackage{amssymb}
\usepackage{hyperref}
\usepackage{dsfont}
\usepackage{bbm}
\usepackage{bm}

\usepackage{gensymb}

\newcommand{\be}[0]{\begin{equation}}
\newcommand{\ee}[0]{\end{equation}}
\newcommand{\ba}[0]{\begin{eqnarray}}
\newcommand{\ea}[0]{\end{eqnarray}}
\newcommand{\mx}[0]{\begin{pmatrix}}
\newcommand{\ex}[0]{\end{pmatrix}}

\newcommand{\abf}{\ensuremath{\bm a}}

\newcommand{\kbf}{\ensuremath{\bm k}}
\newcommand{\qbf}{\ensuremath{\bm q}}
\newcommand{\Qbf}{\ensuremath{\bm Q}}
\newcommand{\Sbf}{\ensuremath{\bm S}}
\newcommand{\Rbf}{\ensuremath{\bm R}}
\newcommand{\rbf}{\ensuremath{\bm r}}

\newcommand{\up}[0]{\uparrow}
\newcommand{\dn}[0]{\downarrow}

\newcommand {\jc}{\textcolor {black}}

\usepackage{stackengine}
\newcommand\xrowht[2][0]{\addstackgap[.5\dimexpr#2\relax]{\vphantom{#1}}}

\title{
Metallic quantum criticality
enabled by flat bands\\ in a kagome lattice
}

\author
{\normalsize{Lei\ Chen$^{1,2}$,  Fang\ Xie$^{1}$,  Shouvik\ Sur$^1$,  Haoyu Hu$^{3}$, }\\
\normalsize{Silke\  Paschen$^{4,1}$, Jennifer Cano$^{2,5}$, Qimiao Si$^{1,\ast}$ }
\\
\small{$^{1}$
Department of Physics and Astronomy, Extreme Quantum Materials Alliance, Smalley-Curl Institute, }\\
\small{Rice University, Houston, Texas, 77005, USA}\\
\small{$^{2}$
Department of Physics and Astronomy, Stony Brook University, Stony Brook, NY 11794, USA }\\
\small{$^{3}$
Donostia International Physics Center, P. Manuel de Lardizabal 4 20018 Donostia-San} \\
\small{{Sebastian, Spain}}\\
\small{$^{4}$
Institute of Solid State Physics, Vienna University of Technology, Wiedner Hauptstr. 8-10, 1040 } \\
\small{Vienna, Austria
}\\
\small{$^{5}$
Center for Computational Quantum Physics, Flatiron Institute, New York, NY 10010, USA }
}

\date{}

\begin{document}


\baselineskip24pt


\maketitle 

\begin{sciabstract}
    Strange metals arise in a variety of platforms for strongly correlated electrons, ranging from the cuprates, heavy fermions to flat band systems. Motivated by recent experiments in kagome metals, we study a Hubbard model on a kagome lattice whose noninteracting limit contains flat bands. A Kondo lattice description is constructed, in which the correlation effects are captured by symmetry preserving and exponentially localized molecular orbitals. These compact molecular orbitals represent the local degrees of freedom that emerge from topological flat bands.
    We identify a quantum critical point at which quasiparticles are
    lost and strange metallicity emerges. Our theoretical work opens up a new route for realizing beyond-Landau quantum criticality, 
    as well as the associated strange metallicity and emergent quantum phases. 
\end{sciabstract}
\vskip 0.1cm

{ \small{$^\ast$To whom correspondence should be addressed; E-mail:  qmsi@rice.edu.} }

\clearpage


\section*{Introduction}

In quantum materials, strong correlations give rise to a rich landscape of quantum 
phases\,\cite{Kei17.1,Pas21.1}.
Microscopically, strong correlations arise when the partially-filled atomic orbitals near the Fermi energy 
make the electrons to experience a larger
electrostatic repulsive interaction  
than their kinetic energy~\cite{Imada1998,Coleman-Schofield05,Si-HF}. The advent of moir\'{e} systems has highlighted the notion of flat bands -- bands of Bloch electron states that hardly disperse~\cite{Bistritzer2011,Cao2018}.
Such flat bands are 
also being investigated in materials with kagome and other line-graph lattices\,\cite{Mielke1991,Checkelsky2018,Yao18.1x,Comin2020-2}, where a
destructive interference of electron motion leads to a reduction in the bandwidth and proportionally enhances the effect of electron interaction. These studies represent a part of extensive ongoing efforts on metals with kagome and related 
crystal structures. Most of the existing efforts  have focused on static 
though unusual electronic orders such as charge density wave\,\cite{Hasan2020,Guguchia2022-TRSB,Zhou2022,Dai2022,Yin2022,Setty2022}.

A central theme of strongly correlated electrons is strange metallicity, which is signified by anomalous temperature and/or frequency dependences in their electrical transport, spin/charge dynamics and thermodynamic properties~\cite{Philip2022,Hu-qcm2022,Kirchner2020,Vojta-RMP}.
Such metals are typically located at the border of electron localization, 
as implicated by 
a jump of the Fermi surface~\cite{paschen2004,Shishido2005,Park2006};
when they are associated with metallic
quantum criticality, they implicate 
a type of quantum critical points (QCPs) that go beyond the Landau framework of order-parameter
fluctuations\,\cite{Hu-qcm2022,Si-Nature,Colemanetal,senthil2004a}.
Recently, strange metallicity 
has also been observed in metals
of kagome and related line-graph 
crystalline structures\,\cite{Huang_pyrochlore_2023.x,Ye2021.x,Ekahana2021.x,Liu-kagome2024}.
raising the question about its origin and, more generally, whether and how it is connected with the strange metallicity of more familiar platforms. In particular, a temperature-pressure phase diagram\,\cite{Liu-kagome2024} in a new kagome metal with active (i.e., proximate-to-the-Fermi-energy) flat band\,\cite{Guo-Yi2024} opens up an outstanding question: How is their strange metallicity linked to quantum criticality in such a topological flat-band setting?

In kagome metals, the destructive interference of electron motion underlies 
{the} flat bands.
Electrons partially occupying a flat band are prone to developing quantum fluctuations that go beyond static ordering. To study such fluctuations,
it is important to treat the correlation effects of the flat bands in terms of variables that preserve the symmetry of the Hamiltonian. The flat bands
can be visualized in terms of compact localized states in real space \cite{Leykam_local-mode2018,Bergman2008}. Consider the case of a kagome metal, a two-dimensional lattice built up from corner-sharing triangles [Fig.\,\ref{fig:illu}A]. 
The wavefunction of the 
compact localized state,
with its amplitude and phase illustrated in Fig.\,\ref{fig:illu}B,
suggests that the flat band could possibly be described by a molecular orbital.
The challenge, however, is that the flat band is topological\,\cite{Tang2011,Ma2020}.
The latter 
prevents a representation of the flat band alone 
in terms of exponentially localized and
{symmetry preserving} Wannier orbitals\,\cite{Soluyanov2011}.
Indeed, the compact localized state illustrated in
Fig.\,\ref{fig:illu}B is not a proper representation of the flat band: such local states from different sites do not form a complete orthonormal basis \cite{Leykam_local-mode2018,Bergman2008}.

Here we 
construct the proper molecular orbitals 
--the compact molecular orbitals (CMOs)-- through a Wannierization procedure that resolves the topological obstruction.
These tight molecular orbitals are represented by the most localized Wannier orbitals
[see Fig.\,\ref{fig:bd}B and the Supplementary Materials (SM), Sec.\,C)],
which predominantly though not exclusively
incorporate the flat bands and
form a triangular lattice 
[Fig.\,\ref{fig:illu}A].
{These orbitals are orthogonal to each other (see the SM, Sec.\,C), in contrast to the compact localized states.}
They are accompanied by  more extended Wannier
orbitals,
which are 
mostly associated with 
the wide bands and  describe the more 
extended molecular orbitals. 
The topological nature of the bands forces a hybridization between the tight and 
extended molecular orbitals.
The use of molecular orbitals is
essential for the range of interactions ($U$) that are large compared to the width of the flat bands ($U>D_{\rm flat}$) and small with respect to the width of the wide bands
($U<D_{\rm wide}$).
Molecular orbitals 
{have}
recently 
{been considered} 
in a much simpler clover lattice,
which
has only a mirror symmetry 
\cite{Hu_fb_2022,Chen_fb_2022}.
Our work 
{for the first time advances the CMOs that connect 
 with the picture of} compact localized states
{while succeeding in resolving all their inconsistencies, both of which are crucial in attaining an intuitively clear real-space representation. In turn, the resulting description by the 
CMOs enables
the realization of}
a theoretical phase diagram [Figs.\,
\ref{fig:illu}B and \ref{fig:phadg}], which provides the first 
explanation of the phase diagram uncovered experimentally\,\cite{Liu-kagome2024}.
Furthermore, we 
demonstrate
{the interacting nature of the QCP}
(Fig.\,\ref{fig:chi})
{and, thus, the strange metallicity} at the 
QCP.
We stress that
kagome lattice has considerably higher symmetry and, indeed, for a long time it has been thought that the construction of exponentially localized and 
{symmetry preserving} Wannier orbitals is not possible~\cite{Ehud2010}. 

From our construction, an 
Anderson/Kondo lattice description arises,
in which
the tight and  
extended molecular orbitals
act as effective atomic-like ($f$) and conduction electron ($c$) degrees of freedom.
{The resulting metallic QCP}
[Fig.\,\ref{fig:illu}D]
features the physics of Kondo destruction viewed from the Anderson/Kondo lattice perspective~\cite{Hu-qcm2022,Si-Nature,Colemanetal,senthil2004a}. Thus, our results reveal a profound 
link of the kagome metals' metallic quantum criticality and concomitant strange metallicity 
with their more established heavy fermion
counterparts\,\cite{Kirchner2020,Hu-qcm2022,Si-Nature,Colemanetal,senthil2004a}.
As such, they
provide an understanding of the strange-metal behavior recently observed in the kagome and related frustrated-lattice systems\,\cite{Huang_pyrochlore_2023.x,Ye2021.x,Ekahana2021.x,Liu-kagome2024}, and uncover a new setting for 
beyond-Landau quantum criticality
and associated emergent quantum phases.

\section*{Results}

\textit{Two-orbital Hubbard model on the kagome lattice:} 
Motivated by the electronic structure 
and physical properties of the
line-graph materials\,\cite{Huang_pyrochlore_2023.x,Ye2021.x,Ekahana2021.x,Liu-kagome2024}, we consider a two-orbital Hubbard model defined on the kagome lattice which follows the crystalline space group (SG) 191. 
The $C_{2z}$ and inversion symmetries
allow for a spin-orbit coupling (SOC) along the $z$ direction only,
which preserves the
$U(1)$ spin-$S_z$ rotational symmetry~\cite{Tang2011}.
The topological flat band, 
possessing 
a nonzero spin Chern number,
is accompanied by a dispersive band above the flat band 
with an opposite spin Chern number. 
The full Hamiltonian is written as $\mathcal{H} = \mathcal{H}_0 + \mathcal{H}_1$, where $\mathcal{H}_0$ is the kinetic 
energy term and $\mathcal{H}_1$ describes the onsite Hubbard interaction.  The band structure of $\mathcal{H}_0$ with the parameters described in the Method is displayed in Fig.\,\ref{fig:bd}A.
Further details of the model are 
provided in the Methods
and in the 
SM (Sec.\,A). 

\textit{Compact molecular orbitals:} 
We proceed with the analysis of the symmetry properties of the Hamiltonian. For a set of bands, if their little-group irreducible representations (irreps) at high symmetry points can be obtained by a combination of orbitals that follows the irreps of the
site symmetry group, 
they are represented by
some elementary band representations (EBRs)\,\cite{Bradlyn2017,Cano2018,Cano2021band}. As such, it becomes possible to establish an ``atomic limit'' that adheres to the crystal symmetry and can be represented by a set of exponentially localized symmetric Wannier functions. In the 
present model, 
considering the presence of SOC, we utilize the double group to describe the symmetry representations. 
Doing so, we find that the wide band just below the middle flat band 
permits, on its own,
Wannierization by symmetric exponentially localized functions. Further, the associated little group irreps for the
middle two bands at high symmetry points ($\Gamma$, $K$, and $M$) are depicted in Fig.\,\ref{fig:bd}B. 
By interchanging the irreps at the $\Gamma$ points of these two bands, they can be transformed into EBRs that correspond to irreps of the site symmetry group at Wyckoff position $1a$ 
(see the Methods and SM, Secs.\,A and \,B, for details).
Consequently, these two bands can be expressed in terms of two well-defined Wannier orbitals, which are symmetric and exponentially-localized and 
which transform under these two specific irreps. A comprehensive analysis of the EBRs is provided in the SM, Sec.\,A.

We use the Wannier90 package~\cite{Pizzi2020} by employing trial orbital wave functions with point group symmetries that satisfy the EBR of the bands of interest to obtain both the maximally-localized Wannier orbitals and the corresponding tight binding parameters 
(see the SM, Sec.\,B, for details). The obtained Wannier functions are depicted in the SM Fig.\,\ref{fig:wn}A,B. The Wannier center
is at the center of the kagome lattice, leading to a tight binding model defined on the triangular lattice as illustrated in 
Fig~\ref{fig:illu}A. Importantly, the flat band is predominantly represented by a single Wannier orbital except in the vicinity of the $\Gamma$ point [see Fig.\,\ref{fig:bd}C]. 
This orbital is the sought-after CMO, which we denote as 
$WN_f$; 
the other, orthogonal and more extended one is called $WN_c$. 
{We stress that the CMO 
$WN_f$
shares the same symmetry --and greatly overlaps-- 
with the compact localized state derived from the flat band wavefunction in the kagome lattice.}
In practice, another dispersive band, which is beneath the targeted flat band, is also close to the flat band. Because this band is topologically trivial, it can
be readily Wannierized on its own. The band structure of the effective three-orbital model is shown in Fig.\,\ref{fig:bd}B. 
At low energies, the system flows to a single channel Kondo fixed point dictated by the conduction electron band that hybridizes more strongly with the $f$ orbitals. 
In our case, the upper conduction band has a stronger hybridization
because the band topology requires the Wannier functions to have weight in both bands,
as  shown in Fig.\,\ref{fig:bd}C.
We refer to the SM (Sec.\,E) for further details. 

\textit{Effective Anderson model via the compact molecular orbitals:}
Importantly, and in contrast to the compact localized states, the
CMOs $WN_f$ from the different unit cells form a complete orthonormal basis (as described in detail in the SM, Sec.\,C). The same is true for the 
orbitals $WN_c$. Accordingly, they represent the proper molecular orbitals.

We proceed to project the Hubbard model of the original lattice in terms of the CMOs.
This results in an effective model expressed as $H_{\text{eff}} = H_0 + H_{\text{int}}$. The kinetic term is described in detail in the Methods section and SM, Sec.\,D.
For the interaction terms, we focus on the dominant on-site Hubbard interaction acting on the 
$f$ electrons. As for the $c$ electrons, their interactions are relatively small compared to their bandwidth and, hence, are inconsequential and can be neglected. 
The final form of the interaction  part
is given by:
\begin{equation}
    H_{int} = \frac{u}{2} \sum_i \left(\sum_{\sigma} f_{i\sigma}^{\dagger} f_{i\sigma} -1 \right)^{2} + \sum_{ij} I_{ij} \Sbf_{i}^{f} \cdot \Sbf_{j}^{f} \, .
\end{equation}
Here, $f^{\dagger}_{i\sigma}$ creates 
an $f$ electron with the $WN_f$ Wannier wavefunction at the unit cell $i$ and  spin $\sigma$, $\Sbf^f=f^{\dagger}\frac{\bm{\sigma}}{2}f$ is the spin operator of the $f$ electron, and $u$ represents the local Hubbard interaction.
The Ruderman–Kittel–Kasuya–Yosida (RKKY) exchange interactions among the $f$ electrons emerges at the low energy limit with $I_{ij}\sim \rho_0J_K^{2}\sim \rho_0 (\frac{4V_{ij}^2}{u})^{2}$, where $V_{ij}$ is the hybridization between the $f$ and $c$ electrons at the positions $i$ and $j$ and $\rho_0$ is the bare density of states of the conduction electrons.
To take into account the dynamical competition between the hybridization and RKKY interactions, we employ the extended dynamical mean-field theory (EDMFT) approach~\cite{Hu-qcm2022,Hu2022.edmft}. In this method, the correlation functions of
the Anderson lattice model 
are calculated in terms of a self-consistently determined Bose-Fermi Anderson (BFA) model~\cite{Hu-qcm2022,zhu2002,Cai_2019,Hu2022.edmft},
which couples the local moment to 
both bosonic and fermionic baths. The self-consistent equations governing the EDMFT calculations are 
described in the Methods.

\begin{figure}[h]
\centering
\includegraphics[width=1\columnwidth]{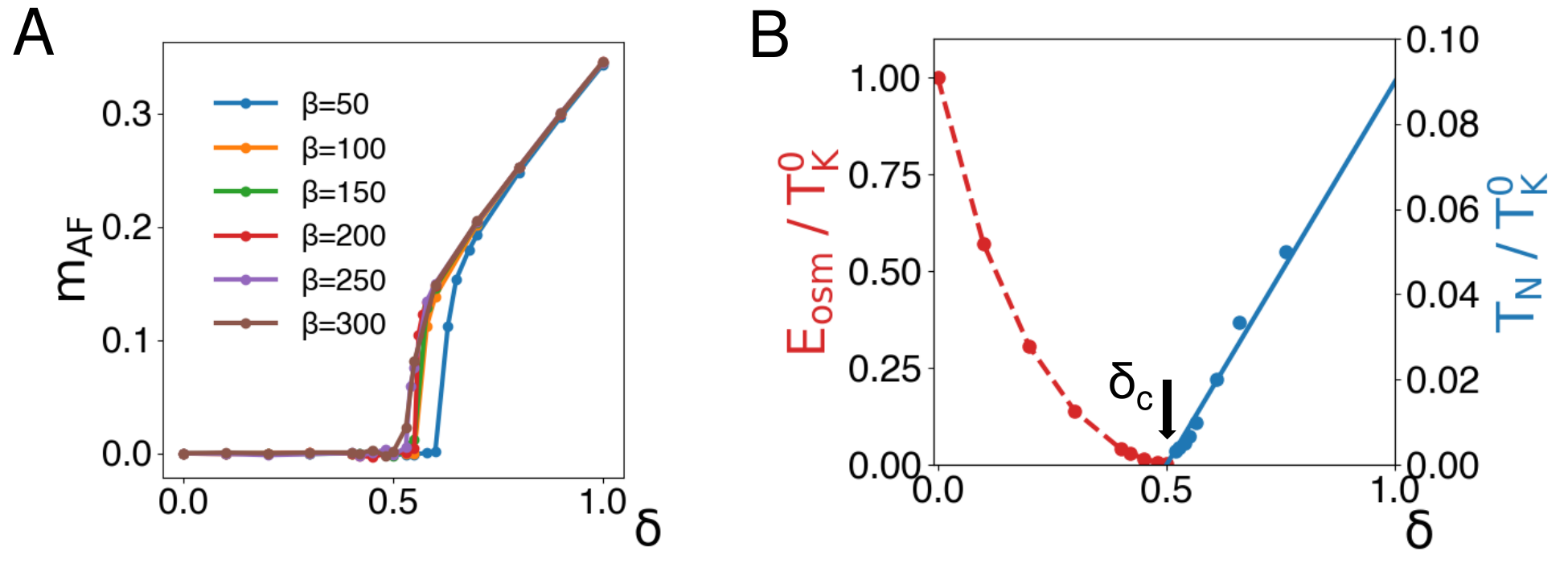}
\caption{ \textbf{Phase diagram and the orbital-seletive Mott energy scale.}
{\bf A}, The amplitude of the antiferromagnetic (AF) order parameter $m_{\rm AF}$ vs. the 
tuning parameter $\delta$, which characterizes the 
dependence 
on the effective Hubbard interaction $u$ of the RKKY interaction to the Kondo scale ratio.
{\bf B}, Phase diagram showing how the N\'eel temperature $T_N$ and the orbital-selective Mott energy scale $E_{\rm osm}$ evolve with $\delta$. 
}
\label{fig:phadg}
\end{figure}

We now determine 
the phase diagram. Without a loss of generality,
we fix $u=0.5$ (in units of the nearest-neighbor intra-orbital hopping parameter of the original Hubbard model)
and take an RKKY density of state of the form $\rho_I(\epsilon)=\Theta(2I-|\epsilon|)/4I$ to incorporate the two-dimensional magnetic fluctuations within a kagome plane, with $4I$ denoting the maximal amplitude of the RKKY interaction in wave vector space. We further define $\delta=I/T_{K}^0$, where $T_{K}^0$ is the underlying Kondo energy scale at $I=0$. Because $\delta$ is a function of $u$, tuning $\delta$ is tantamount to 
varying the 
interaction $U$ of the original Hubbard model
(Eq.\,\ref{Hubbard_original}).
We perform the calculations 
at various temperatures to scan the phase diagram. As illustrated in Fig.\,\ref{fig:phadg}A,
for a fixed temperature, a 
magnetic 
phase transition is indicated by the onset of the order parameter ($m_{AF}$) 
upon increasing $\delta$. The incipient jump of the order parameter decreases with the lowering of temperature and is extrapolated to zero in the zero-temperature limit~\cite{Hu2022.edmft}, which
corresponds to a continuous 
quantum phase transition. 
Besides the 
suppression of magnetic order,
the Anderson lattice model also undergoes a 
dehybridization 
between the 
$f$ and 
$c$ electrons, 
which is characterized by the
``orbital-selective Mott" energy scale 
$E_{\rm osm}$. The 
phase diagram is shown in Fig.\,\ref{fig:phadg}. 
One can 
observe that the 
N\'eel temperature develops 
at the same point 
where the 
energy scale, $E_{\rm osm}$, goes to zero.
From the Kondo 
perspective, 
the transition corresponds to a Kondo-destruction 
QCP.
In terms of the original Hubbard model, this is a continuous selective Mott transition of the molecular orbitals.

We next turn to studying the nature of the
QCP with a special focus on its dynamical properties.
We calculate both the local and lattice spin susceptibilities, which are denoted as $\chi_{loc}(i\omega_n)$ and $\chi(\Qbf, i\omega_n)$, respectively (see 
the Methods). As shown in Fig.\,\ref{fig:chi}A, at the QCP, $\chi_{loc}(i\omega_n)$ follows a logarithmic divergence in the low frequency domain. 
The lattice spin susceptibility 
at the magnetic ordering wavevector $\Qbf$ 
is shown in Fig.\,\ref{fig:chi}B:
It obeys an $\hbar \omega/k_{\rm B} T$ scaling in the quantum critical region, signifying that the 
QCP is interacting (as opposed to Gaussian) and $k_{\rm B}T$ is the universal and only energy scale describing the quantum criticality~\cite{Hu-qcm2022}. The 
critical exponent for the dynamical spin susceptibility is fitted to be 
$\alpha \approx 0.8$; the fact that it is fractional further underscores the interacting nature of the fixed point. 
 
\section*{Discussion}
{Our work is unprecedented in that i) we have resolved the long-standing inconsistency of the compact localized states for kagome metals; 
ii) using the resulting CMO-based effective model,
we have realized a phase diagram for the kagome metals, in which a non-thermal control parameter tunes the system from a static electronic order to a paramagnetic metal through a
QCP; and iii) we have found a dynamical Planckian 
($\hbar \omega/k_{\rm B}T$) scaling 
 at the 
 QCP that is
enabled by the kagome lattice's flat bands,
via the CMOs.
These
properties 
uncover a striking degree of universality between the flat-band-active kagome metals and} the well established strange metallicity 
of heavy fermion systems.
The $\hbar \omega/k_{\rm B}T$ scaling implicates a linear-in-$T$ relaxation rate, which provides 
the understanding of the puzzling observations of the linear-in-energy damping rate in the ARPES spectrum and linear-in-$T$ resistivity in 
the aforementioned kagome materials\,\cite{Ye2021.x,Ekahana2021.x}.
As such, our theoretical phase diagram explains the salient features 
of the striking experimental observations in a new kagome metal with active flat bands\,\cite{Liu-kagome2024}.
The dynamical scaling can be tested in future measurements of the inelastic neutron scattering cross section
and optical conductivity in these systems. 
Finally, the selective transition of the molecular orbitals
has a salient signature in a jump of Fermi surface as the system goes across the 
QCP 
on the tuning axis $\delta$. 
This theoretical expectation can be tested by
ARPES, Hall effect and quantum oscillation measurements in the aforementioned line-graph materials 
in the low-temperature regime (once the experimental tuning 
axes are established that go across the QCP),
by analogy with the well-established Hall and quantum oscillation measurements in quantum critical heavy fermion metals~\cite{paschen2004,Shishido2005,Park2006};
in particular, we propose to carry out such measurements across the pressure-induced QCP that has been observed in Ref.\,\cite{Liu-kagome2024}.
The quantum criticality we have uncovered is 
closely related to the
Kondo-destruction quantum criticality.
As such,
the kagome  and related-frustrated-lattice metals with active flat bands represent a new platform for realizing and probing 
{the} beyond-Landau quantum criticality\,
\cite{Hu-qcm2022,Kirchner2020,paschen2004,Shishido2005,Park2006,Si-Nature,Colemanetal,senthil2004a}.

The orbital-selective Mott criticality is also expected to promote emergent quantum phases.
A promising emergent phase is 
superconductivity.
By analogy with the case of heavy-fermion quantum criticality\,\cite{Hu2021-SC}, we can expect
the superconductivity to have 
a high transition temperature (as measured by the natural underlying Fermi temperature scale).
Our work, thus,
raises the prospect for the development of unconventional superconductivity in kagome metals\,\cite{Liu-kagome2024}.

Going beyond kagome metals, 
our results make it natural for the flat bands in other line-graph settings, such as the pyrochlore lattice, to enable 
a similar form of quantum criticality. In turn, this provides 
an understanding of the recent experimental
observation of strange metallicity in a pyrochlore metal\,\cite{Huang_pyrochlore_2023.x}. 
More generally, our work showcases the link between the transition-metal line-graph metals, heavy fermions and 
moir\'{e} systems, where a 
Kondo perspective has also been theoretically fruitful\,\cite{Ram2021,Song2022,Kumar2022,Guerci2022.x,Shan2023} and where 
strange-metal behavior is being experimentally uncovered\,\cite{Cao20-PRL,Ghiotto2021,Jao22}.

In conclusion, we have advanced a realistic model study  on the effect of local Coulomb interactions 
in the flat bands of a kagome metal.
We have constructed the compact molecular orbitals,
which, in contrast to the
compact localized 
states, form a proper basis. This leads
to an Anderson/Kondo lattice description for the selective correlations of the molecular orbitals,
which in turn allows
a non-perturbative study that uncovers a continuous selective transition of the molecular orbitals. The resulting quantum critical metal has
the hallmarks of strange metallicity. 
Our work allows the first theoretical
understanding of the emerging experiments on strange metal behavior in line-graph materials,
and answers the aforementioned question raised by a striking new 
experimental phase diagram\,\cite{Liu-kagome2024} for kagome metals with active flat bands.
Furthermore, our work predicts the properties accompanying the strange metallicity 
in single-electron excitations and collective dynamics, and 
raises the prospect for unconventional superconductivity in these systems.
Our findings also reveal new interconnections among a variety of correlated electron platforms, and point to new platforms for beyond-Landau quantum criticality.

{\bf Note added:~}Since the circulation of this manuscript,
the experimental work of J. C. Souza et al. has used STM measurements to 
observe the signatures of the CMOs advanced here.

\section*{Methods}
\textit{Hubbard model on the kagome lattice:} The
kagome lattice contains three sublattices. 
The site symmetry group for each sublattice is $D_{2h}$, and the local orbitals can be classified into orbitals of even and odd according to the eigenvalues of the $C_{2z}$ symmetry (see the details in the SM, Sec.\,A). 
For the physical $3d$-orbitals of the transition-metal systems, there are two classes according
to the $C_{2z}$ symmetry:
The $3d_{z^2}/3d_{xy}/3d_{x^2-y^2}$ orbitals are even, while the $3d_{xz}/3d_{yz}$ orbitals are odd under $C_{2z}$.
We consider a two-orbital model defined on the kagome lattice, with one even and one odd $3d$ orbital denoted as $\psi_1$ and $\psi_2$ respectively. It is written as $\mathcal{H} = \mathcal{H}_0 + \mathcal{H}_1$, which contains the on-site Hubbard interaction,
\begin{equation}
    \mathcal{H}_1 = \frac{U}{2} \sum_{i,\eta} \left ( \sum_{\tau\sigma} n_{i,\eta,\tau,\sigma} - 2 \right)^2 
\label{Hubbard_original}
\end{equation}
where $\tau = 1, 2$ and $\sigma = (\up, \dn)$ denote the orbital and spin indices, respectively, and $\eta = A/B/C$ goes through the three sublattices in a unit cell. The kinetic Hamiltonian is given by $\mathcal{H}_0 = \sum_{\kbf \sigma}\Psi^{\dagger}_{\kbf \sigma} \mathcal{H}_0^{\sigma}(\kbf) \Psi_{\kbf \sigma} $, where $\Psi_{\kbf \sigma} = [\psi_{A\tau \sigma}, \psi_{B\tau \sigma}, \psi_{C\tau \sigma}]^{T} $. 
As described earlier, there only exists
spin-orbital coupling along the $z$ direction; thus,$\mathcal{H}_0^{\up} (\kbf) = [\mathcal{H}_0^{\dn} (\kbf)]^*$~\cite{Tang2011} and 
\begin{equation}
    \mathcal{H}_0^{\up}(\kbf) = \mx
    \mathcal{H}^{11}_0(\kbf) & \mathcal{H}_0^{12}(\kbf) \\
    [\mathcal{H}_0^{12}(\kbf)]^{\dagger} & \mathcal{H}_0^{22}(\kbf)
    \ex,
\end{equation}
where 
\begin{equation}
    \mathcal{H}_0^{\tau\tau}(\kbf) = \mx
    -\epsilon_{\tau} & 2(- t_{\tau} +i\lambda_{\tau}) \cos k_1  & 2(- t_{\tau} -i\lambda_{\tau}) \cos k_2 \\
    2(- t_{\tau} - i\lambda_{\tau}) \cos k_1 & -\epsilon_{\tau} & 2(- t_{\tau} + i\lambda_{\tau}) \cos k_3 \\
    2(- t_{\tau} + i\lambda_{\tau}) \cos k_2 & 2(- t_{\tau} - i\lambda_{\tau}) \cos k_3 & -\epsilon_{\tau}
    \ex,
\end{equation}
and 
\begin{equation}
    \mathcal{H}_0^{12}(\kbf) = \mx
    0 & -2i t_{12} \sin k_1 & 2 i t_{21} \sin k_2 \\
    -2i t_{21} \sin k_1 & 0 & -2 i t_{12}  \sin k_3 \\
    2 i t_{12} \sin k_2 & -2 i t_{21}  \sin k_3  & 0
    \ex.
\end{equation}
The 
basis vectors of the kagome lattice are $\bm{a}_{1/2}=\frac{1}{2} \bm{x} \pm \frac{\sqrt{3}}{2}\bm{y}$. In addition, $\bm{d}_1=\frac{1}{2}(\bm{a}_1 + \bm{a}_2) $, $\bm{d}_2=\frac{1}{2}\bm{a}_1$, $\bm{d}_3=-\frac{1}{2}\bm{a}_2$ are the displacements between 
the different sublattices, and $k_i = \kbf \cdot \bm{d}_{i}$ [Fig.\,\ref{fig:org}B and C]. 
Furthermore, $t_{\tau}$ and $\lambda_{\tau}$ denote the nearest-neighbor intraorbital hopping and spin-orbital coupling respectively, $\epsilon_{\tau}$ is the energy level for the orbital $\tau$, and $t_{12}/t_{21}$ are the symmetry-allowed interorbital hoppings. Without a loss of generality, we adapt the following parameter setting with
$t_1=t_2=1$, $\lambda_1=\lambda_2=-0.18$, $\epsilon_1=-2.75$, $\epsilon_2=4.25$ and $t_{12}=t_{21}=-0.1$.
A detailed analysis of the dispersion and symmetry is provided in the SM, Sec.\,A.

\textit{Effective extended Anderson model via the CMOs:}
We focus on the flat and dispersive bands close to the Fermi energy. 
By interchanging the irreps at the $\Gamma$ points of these two bands, we find that they develop into the induced representations $\bar{E}_{1g}\up G$ and $\bar{E}_{1u}\up G$~\cite{Group}, where $\bar{E}_{1g}$ and $\bar{E}_{1u}$ are the site symmetry group irreps at Wyckoff position $1a$.
In other words, these two bands can be represented by a set of exponentially localized Wannier orbitals, which transform as these site symmetry group irreps, without topological obstruction (See the SM, Secs.\,A and \,B). 
The Wannier centers of the emergent orbitals turn out to form a triangular lattice. The flat band is mostly captured by one orbital, which we call $WN_{f}$, while the dispersive band is mostly represented by the other one which we name $WN_{c}$. We project the Hubbard model of the original lattice to the Wannier basis. This leads to an effective Anderson model expressed in terms of the $f$ and $c$ electrons with $H_{eff}=H_0 + H_{int}$. Here, the kinetic term is:
\begin{equation}
\begin{aligned}
    H_0 & = H_{f} + H_{c} + H_{v} \\
    & = \sum_{ij, \sigma} t_{ij}^{f}\left( f_{i\sigma}^{\dagger} f_{j\sigma} + h.c. \right) - \sum_{i} \epsilon_f f_{i\sigma}^{\dagger} f_{i\sigma} \\
    & +  \sum_{ij,\sigma} t_{ij}^c \left( c_{i\sigma}^{\dagger} c_{j\sigma} + h.c. \right) - \sum_{i\sigma} \epsilon_c c_{i\sigma}^{\dagger} c_{i\sigma} \\
    & + \sum_{ij\sigma} \left( V^{\sigma}_{ij}  f_{i\sigma}^{\dagger} c_{j\sigma} +h.c.\right) \, .
\end{aligned}
\end{equation}
Here, $f_{i\sigma}^{\dagger}$ ($c_{i\sigma}^{\dagger}$) creates a heavy (light) electron at the position $i$ with spin $\sigma$, $t^{f}_{ij}$ ($t_{ij}^{c}$) describes the hopping 
among the $f$ ($c$) electrons between the positions $i$ and $j$, and the $|t^f|$'s, the hopping amplitudes between the $f$ orbitals, are much smaller than the $|t^c|$'s, the hopping amplitudes between the $c$ orbitals. In addition, $V^{\sigma}_{ij}$ represents the hybridization between the light and heavy orbitals. According to time reversal symmetry, $V^{\up}_{ij} =(V^{\dn}_{ij})^{*}$. Finally, $\epsilon_c$ ($\epsilon_f$) denotes the energy level of the $c$ ($f$) orbital in the noninteracting limit.  

\textit{Extended dynamical mean field theory and self-consistent equations:} The EDMFT method can be used to 
analyze the dynamical competition between the hybridization and RKKY interactions. We treat the model $H_{eff}$ with EDMFT by calculating the correlation functions of the lattice model in terms of an effective Bose-Fermi Anderson (BFA)
model:
\begin{equation}
\begin{aligned}
      S_{BFA} &= \sum_{\omega, \sigma, a, b} \phi^{\dagger}_{a,\sigma} [(-i\omega_n - \epsilon)\delta_{ab} + \Delta_{ab}(i\omega_n) ] \phi_{b,\sigma} \\
      & - \frac{1}{2}\sum_{\Omega,\nu=x,y,z} S^{\nu} (i\Omega_n) \chi_0^{\nu,-1}(i\Omega_n) S^{\nu} (-i\Omega) +\int_{\tau} d \tau \,h_0\,,
\end{aligned}
\end{equation}
where $\epsilon = \mathrm{diag}(\epsilon_f, \epsilon_c)$ is the energy levels written in a matrix form. 
In the BFA model, the local orbitals are coupled to both the fermionic baths [$\Delta(i\omega_n)$] and the bosonic bath [$\chi_0^{\mu}[i\Omega_n)$], which capture the dynamics of the
hybridization and spin correlations, respectively. The bath functions $\Delta(i\omega_n)$ and $\chi_0^{\nu,-1}$ are determined 
via the following self-consistent equations:
\begin{equation}\label{eq:self}
\begin{aligned}
        G_{loc} (i\omega_n) &= \sum_{\kbf} [i\omega_n - H_0(\kbf) - \Sigma(i\omega_n)]^{-1} \\
        & = [(i\omega_n + \epsilon - \Delta(i\omega_n))^{-1} - \Sigma(i\omega_n)]^{-1}, \\
        \chi_{loc}^{\nu} (i\Omega_n) &= \sum_{\qbf} [J_{\qbf}^{\nu} + M^{\nu}(i\Omega_n)]^{-1} \\
        & = [\chi_{0}^{\nu,-1}(i\Omega) + M^{\nu}(i\Omega_n)]^{-1}.
\end{aligned}
\end{equation}
Here, $G_{loc}(i\omega_n)$ and $\chi_{loc}^{\nu}(i\Omega_n) $ are the single-particle Green's function of the local cluster and the local spin-spin correlation functions, $\Sigma(i\omega_n)$ and $M^{\nu}(i\Omega_n)$ are the self-energy and spin cumulant, respectively. Because the $f$ and $c$ orbitals belong to different irreducible representations, the on-site hopping between the orbitals is forbidden. Therefore $G_{loc}(i\omega_n)$, $\Sigma(i\omega_n)$ and $\Delta(i\omega_n)$ are block diagonal~\cite{Kotlier2022,Hu2022-orbital}. As the spatial spin interactions of the 
$f$ orbitals are pre-dominant and are considered, only the $f$ orbital is coupled with the bosonic bath in the BFA. 
The existence of the SOC induces spin anisotropy. 
The precise form of spin anisotropy is unimportant, recognizing 
that the emergence of new fixed points in the Bose-Fermi Anderson/Kondo model is insensitive to the spin symmetry~\cite{Hu-qcm2022,zhu2002}.
For simplicity, we 
consider the spatial spin interactions to be Ising anisotropic, with the RKKY interaction to be 
$I \sum_{<ij>}S_{f,i}^{z} S_{f,j}^{z}\,$.

\begin{figure}[h]
\centering
\includegraphics[width=1\columnwidth]{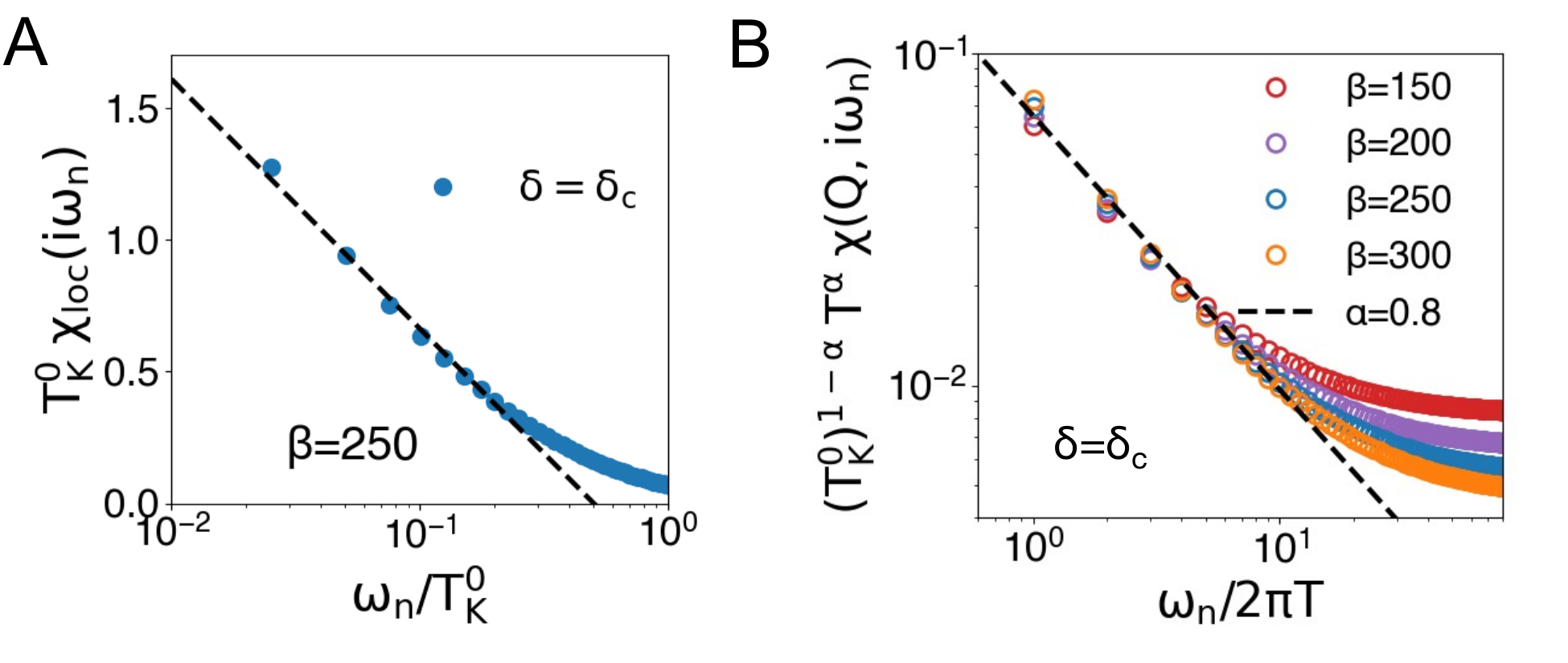}
\caption{ \textbf{Dynamical scaling at the metallic quantum critical point.} {\bf A,} The dynamical local spin susceptibility at the QCP ($\delta=\delta_c$). {\bf B,} $\hbar \omega/k_{\rm B} T$ scaling for the dynamical lattice spin susceptibility, also at the QCP. Here, in the figure legends, $\hbar$ and $k_{\rm B}$ have been set to $1$.} 
\label{fig:chi}
\end{figure}

\textit{Local and lattice dynamical spin susceptibilities:} We first define the dynamical local spin susceptibility as
\begin{equation}
    \chi_{loc}^{\nu}(i\omega_n, T) = \int_{0}^{\beta} d\tau e^{i\omega_n \tau} \langle S^{\nu} (\tau)S^{\nu}(0) \rangle,
\end{equation}
where $S$ is the spin of the impurity model with the spin direction $\nu=x,y,z$.  As shown in Fig.\,\ref{fig:chi}A, at the OSM QCP, the isothermal local spin susceptibility follows a logarithmic function as 
\begin{equation}
    \chi_{loc}(i\omega_n) = \frac{-\alpha_{\omega}}{4I} \log \omega_n.
\end{equation}

The lattice spin susceptibility is determined by the Dyson equation: 
\begin{equation}
    \chi^{\alpha}(\qbf, i\omega) = \frac{1}{M^{\alpha}(i\omega) + I_{\qbf}},
\end{equation}
with $\alpha \in {x,y,z}$. At the magnetic wave vector $\Qbf$, we have 
\begin{equation}
    \chi^{\alpha}(\Qbf, i\omega_n) = \frac{1}{M^{\alpha}(i\omega_n) - 2I}.
\end{equation}
Combining this equation with the self-consistent equations in Eq.~\ref{eq:self}, we have the following asymptotic 
form when $4I\chi_{loc} \gg 1$,
\begin{equation}
    \chi^{\alpha}(\Qbf, i\omega) \approx \frac{\exp(4I\chi_{loc}^{\alpha}(i\omega_n))}{4I}.
\end{equation} 

\noindent
{\bf Data availability:}~~
All data needed to evaluate the conclusions in the paper are presented in the paper and/or
the Supplementary Materials. Additional data that have been used
are available from the corresponding author.

\bibliographystyle{Science}
\bibliography{qcp_ft}

\begin{thebibliography}{10}

\bibitem{Kei17.1}
B.~Keimer, J.~E. Moore, {The physics of quantum materials}, {\it {Nat.\
  Phys.}\/} {\bf 13}, 1045 (2017).

\bibitem{Pas21.1}
S.~Paschen, Q.~Si, Quantum phases driven by strong correlations, {\it Nat. Rev.
  Phys.\/} {\bf 3}, 9 (2021).

\bibitem{Imada1998}
M.~Imada, A.~Fujimori, Y.~Tokura, Metal-insulator transitions, {\it Rev. Mod.
  Phys.\/} {\bf 70}, 1039--1263 (1998).

\bibitem{Coleman-Schofield05}
P.~Coleman, A.~J. Schofield, {Quantum criticality}, {\it {Nature}\/} {\bf 433},
  226 (2005).

\bibitem{Si-HF}
Q.~Si, F.~Steglich, Heavy fermions and quantum phase transitions, {\it
  Science\/} {\bf 329}, 1161-1166 (2010).

\bibitem{Bistritzer2011}
R.~Bistritzer, A.~H. MacDonald, Moiré bands in twisted double-layer graphene,
  {\it Proc. Natl. Acad. Sci. U.S.A.\/} {\bf 108}, 12233-12237 (2011).

\bibitem{Cao2018}
Y.~Cao, {\it et~al.\/}, Unconventional superconductivity in magic-angle
  graphene superlattices, {\it Nature\/} {\bf 556}, 43-50 (2018).

\bibitem{Mielke1991}
A.~Mielke, Ferromagnetic ground states for the {H}ubbard model on line graphs,
  {\it J. Phys. A: Math. Gen.\/} {\bf 24}, L73 (1991).

\bibitem{Checkelsky2018}
L.~Ye, {\it et~al.\/}, Massive {D}irac fermions in a ferromagnetic kagome
  metal, {\it Nature\/} {\bf 555}, 638--642 (2018).

\bibitem{Yao18.1x}
M.~Yao, {\it et~al.\/}, {Switchable Weyl nodes in topological kagome
  ferromagnet Fe$_3$Sn$_2$}, {\it arXiv:1810.01514\/}  (2018).

\bibitem{Comin2020-2}
M.~Kang, {\it et~al.\/}, Topological flat bands in frustrated kagome lattice
  {C}o{S}n, {\it Nat. Commun.\/} {\bf 11}, 1--9 (2020).

\bibitem{Hasan2020}
Y.-X. Jiang, {\it et~al.\/}, Unconventional chiral charge order in kagome
  superconductor {KV$_3$Sb$_5$}, {\it Nat. Mater.\/} {\bf 20}, 1353 (2021).

\bibitem{Guguchia2022-TRSB}
C.~Mielke, {\it et~al.\/}, Time-reversal symmetry-breaking charge order in a
  correlated kagome superconductor, {\it Nature\/} {\bf 602}, 245-250 (2022).

\bibitem{Zhou2022}
S.~Zhou, Z.~Wang, Chern {F}ermi pocket, topological pair density wave, and
  charge-4e and charge-6e superconductivity in kagome superconductors, {\it
  Nat. Commun.\/} {\bf 13}, 7288 (2022).

\bibitem{Dai2022}
X.~Teng, {\it et~al.\/}, Discovery of charge density wave in a kagome lattice
  antiferromagnet, {\it Nature\/} {\bf 609}, 490-495 (2022).

\bibitem{Yin2022}
J.-X. Yin, {\it et~al.\/}, Discovery of charge order and corresponding edge
  state in kagome magnet {FeGe}, {\it Phys. Rev. Lett.\/} {\bf 129}, 166401
  (2022).

\bibitem{Setty2022}
C.~Setty, {\it et~al.\/}, Electron correlations and charge density wave in the
  topological kagome metal {FeGe}, {\it arXiv preprint arXiv:2203.01930\/}
  (2022).

\bibitem{Philip2022}
P.~W. Phillips, N.~E. Hussey, P.~Abbamonte, Stranger than metals, {\it
  Science\/} {\bf 377}, eabh4273 (2022).

\bibitem{Hu-qcm2022}
H.~Hu, L.~Chen, Q.~Si, Quantum critical metals and loss of quasiparticles, {\it
  Nat. Phys.\/} {\bf 20}, 1863-1873 (2024).

\bibitem{Kirchner2020}
S.~Kirchner, {\it et~al.\/}, Colloquium: Heavy-electron quantum criticality and
  single-particle spectroscopy, {\it Rev. Mod. Phys.\/} {\bf 92}, 011002
  (2020).

\bibitem{Vojta-RMP}
H.~v. L\"ohneysen, A.~Rosch, M.~Vojta, P.~W\"olfle, Fermi-liquid instabilities
  at magnetic quantum phase transitions, {\it Rev. Mod. Phys.\/} {\bf 79},
  1015--1075 (2007).

\bibitem{paschen2004}
S.~Paschen, {\it et~al.\/}, Hall-effect evolution across a heavy-fermion
  quantum critical point, {\it Nature\/} {\bf 432}, 881 (2004).

\bibitem{Shishido2005}
H.~Shishido, R.~Settai, H.~Harima, Y.~Ōnuki, A drastic change of the {Fermi}
  surface at a critical pressure in {CeRhIn$_5$}: {dHvA} study under pressure,
  {\it J. Phys. Soc. Jpn.\/} {\bf 74}, 1103-1106 (2005).

\bibitem{Park2006}
T.~Park, {\it et~al.\/}, Hidden magnetism and quantum criticality in the heavy
  fermion superconductor {CeRhIn$_5$}, {\it Nature\/} {\bf 440}, 65-68 (2006).

\bibitem{Si-Nature}
Q.~Si, S.~Rabello, K.~Ingersent, J.~L. Smith, Locally critical quantum phase
  transitions in strongly correlated metals, {\it Nature\/} {\bf 413}, 804-808
  (2001).

\bibitem{Colemanetal}
P.~Coleman, C.~P\'{e}pin, Q.~Si, R.~Ramazashvili, How do {F}ermi liquids get
  heavy and die?, {\it J.~Phys.~Cond.~Matt.\/} {\bf 13}, R723 (2001).

\bibitem{senthil2004a}
T.~Senthil, M.~Vojta, S.~Sachdev, Weak magnetism and non-fermi liquids near
  heavy-fermion critical points, {\it Phys.~Rev.~B\/} {\bf 69}, 035111 (2004).

\bibitem{Huang_pyrochlore_2023.x}
J.~Huang, {\it et~al.\/}, Non-fermi liquid behaviour in a correlated flat-band
  pyrochlore lattice, {\it Nat. Phys.\/} {\bf 20}, 603-609 (2024).

\bibitem{Ye2021.x}
L.~Ye, {\it et~al.\/}, Hopping frustration-induced flat band and strange
  metallicity in a kagome metal, {\it Nat. Phys.\/} {\bf 20}, 610-614 (2024).

\bibitem{Ekahana2021.x}
S.~A. Ekahana, {\it et~al.\/}, Anomalous electrons in a metallic kagome
  ferromagnet, {\it Nature\/} {\bf 627}, 67-72 (2024).

\bibitem{Liu-kagome2024}
Y.~Liu, {\it et~al.\/}, Superconductivity under pressure in a chromium-based
  kagome metal, {\it Nature\/} {\bf 632}, 1032-1037 (2024).

\bibitem{Guo-Yi2024}
Y.~Guo, {\it et~al.\/}, Ubiquitous flat bands in a cr-based kagome
  superconductor, {\it arXiv preprint arXiv:2406.05293\/}  (2024).

\bibitem{Leykam_local-mode2018}
D.~Leykam, A.~Andreanov, S.~Flach, Artificial flat band systems: from lattice
  models to experiments, {\it Adv. Phys.: X\/} {\bf 3}, 1473052 (2018).

\bibitem{Bergman2008}
D.~L. Bergman, C.~Wu, L.~Balents, Band touching from real-space topology in
  frustrated hopping models, {\it Phys. Rev. B\/} {\bf 78}, 125104 (2008).

\bibitem{Tang2011}
E.~Tang, J.-W. Mei, X.-G. Wen, High-temperature fractional quantum hall states,
  {\it Phys. Rev. Lett.\/} {\bf 106}, 236802 (2011).

\bibitem{Ma2020}
D.-S. Ma, {\it et~al.\/}, Spin-orbit-induced topological flat bands in line and
  split graphs of bipartite lattices, {\it Phys. Rev. Lett.\/} {\bf 125},
  266403 (2020).

\bibitem{Soluyanov2011}
A.~A. Soluyanov, D.~Vanderbilt, Wannier representation of ${Z}_{2}$ topological
  insulators, {\it Phys. Rev. B\/} {\bf 83}, 035108 (2011).

\bibitem{Hu_fb_2022}
H.~Hu, Q.~Si, {Coupled topological flat and wide bands: Quasiparticle formation
  and destruction}, {\it Sci. Adv.\/} {\bf 9}, eadg0028 (2023).

\bibitem{Chen_fb_2022}
L.~Chen, {\it et~al.\/}, Emergent flat band and topological kondo semimetal
  driven by orbital-selective correlations, {\it Nat. Commun.\/} {\bf 15}, 5242
  (2024).

\bibitem{Ehud2010}
S.~D. Huber, E.~Altman, Bose condensation in flat bands, {\it Phys. Rev. B\/}
  {\bf 82}, 184502 (2010).

\bibitem{Bradlyn2017}
B.~Bradlyn, {\it et~al.\/}, Topological quantum chemistry, {\it Nature\/} {\bf
  547}, 298-305 (2017).

\bibitem{Cano2018}
J.~Cano, {\it et~al.\/}, Building blocks of topological quantum chemistry:
  Elementary band representations, {\it Phys. Rev. B\/} {\bf 97}, 035139
  (2018).

\bibitem{Cano2021band}
J.~Cano, B.~Bradlyn, Band representations and topological quantum chemistry,
  {\it Annu.\ Rev.\ Condens.\ Matter Phys.\/} {\bf 12}, 225--246 (2021).

\bibitem{Pizzi2020}
G.~Pizzi, {\it et~al.\/}, Wannier90 as a community code: new features and
  applications, {\it J. Phys. Condens. Matter.\/} {\bf 32}, 165902 (2020).

\bibitem{Hu2022.edmft}
H.~Hu, L.~Chen, Q.~Si, Extended dynamical mean field theory for correlated
  electron models, {\it arXiv preprint arXiv:2210.14197\/}  (2022).

\bibitem{zhu2002}
L.~Zhu, Q.~Si, Critical local-moment fluctuations in the {Bose-Fermi Kondo}
  model, {\it Phys. Rev. B\/} {\bf 66}, 024426 (2002).

\bibitem{Cai_2019}
A.~Cai, Q.~Si, {Bose-Fermi Anderson model with SU(2) symmetry: Continuous-time
  quantum Monte Carlo study}, {\it Phys. Rev. B\/} {\bf 100}, 014439 (2019).

\bibitem{Hu2021-SC}
H.~Hu, {\it et~al.\/}, Unconventional superconductivity from {Fermi} surface
  fluctuations in strongly correlated metals, {\it arXiv preprint
  arXiv:2109.13224\/}  (2021).

\bibitem{Ram2021}
A.~Ramires, J.~L. Lado, Emulating heavy fermions in twisted trilayer graphene,
  {\it Phys. Rev. Lett.\/} {\bf 127}, 026401 (2021).

\bibitem{Song2022}
Z.-D. Song, B.~A. Bernevig, Magic-angle twisted bilayer graphene as a
  topological heavy fermion problem, {\it Phys. Rev. Lett.\/} {\bf 129}, 047601
  (2022).

\bibitem{Kumar2022}
A.~Kumar, N.~C. Hu, A.~H. MacDonald, A.~C. Potter, Gate-tunable heavy fermion
  quantum criticality in a moir\'e {Kondo} lattice, {\it Phys. Rev. B\/} {\bf
  106}, L041116 (2022).

\bibitem{Guerci2022.x}
D.~Guerci, {\it et~al.\/}, Chiral kondo lattice in doped
  $\mathrm{MoTe_2}$/$\mathrm{WSe_2}$ bilayers, {\it Sci. Adv.\/} {\bf 9},
  eade7701 (2023).

\bibitem{Shan2023}
W.~Zhao, {\it et~al.\/}, Gate-tunable heavy fermions in a moir{\'e} {Kondo}
  lattice, {\it Nature\/} {\bf 616}, 61-65 (2023).

\bibitem{Cao20-PRL}
Y.~Cao, {\it et~al.\/}, Strange metal in magic-angle graphene with near
  planckian dissipation, {\it Phys. Rev. Lett.\/} {\bf 124}, 076801 (2020).

\bibitem{Ghiotto2021}
A.~Ghiotto, {\it et~al.\/}, Quantum criticality in twisted transition metal
  dichalcogenides, {\it Nature\/} {\bf 597}, 345-349 (2021).

\bibitem{Jao22}
A.~Jaoui, {\it et~al.\/}, {Quantum critical behaviour in magic-angle twisted
  bilayer graphene}, {\it {Nat.\ Phys.}\/} {\bf 18}, 633--638 (2022).

\bibitem{Group}
T.~Inui, Y.~Tanabe, Y.~Onodera, {\it Group theory and its applications in
  physics\/}, vol.~78 (2012).

\bibitem{Kotlier2022}
F.~B. Kugler, G.~Kotliar, Is the orbital-selective mott phase stable against
  interorbital hopping?, {\it Phys. Rev. Lett.\/} {\bf 129}, 096403 (2022).

\bibitem{Hu2022-orbital}
H.~Hu, L.~Chen, J.-X. Zhu, R.~Yu, Q.~Si, Orbital-selective {M}ott phase as a
  dehybridization fixed point, {\it arXiv preprint arXiv:2203.06140\/}  (2022).

\bibitem{BCS}
M.~I. Aroyo, {\it et~al.\/}, Crystallography online: Bilbao crystallographic
  server, {\it Bulg. Chem. Commun\/} {\bf 43}, 183--197 (2011).

\bibitem{Vanderbilt2012-RMP}
N.~Marzari, A.~A. Mostofi, J.~R. Yates, I.~Souza, D.~Vanderbilt, {Maximally
  localized Wannier functions: Theory and applications}, {\it Rev. Mod.
  Phys.\/} {\bf 84}, 1419--1475 (2012).

\end{thebibliography}

\clearpage

\noindent{\bf Acknowledgment:}~~
We thank Gabriel Aeppli, Joseph Checkelsky, Pengcheng Dai, and Ming Yi for useful
discussions. Work at Rice has primarily been supported by the 
National Science Foundation
under Grant No. DMR-2220603
(Conceptualization and Wannier construction, L.C. and F.X.), 
by the Air Force Office of Scientific Research under Grant No.
FA9550-21-1-0356 (Conceptualization and orbital-selective transition, L.C., F.X., S.S.), 
and by the Robert A. Welch Foundation Grant No. C-1411
and the Vannevar Bush Faculty Fellowship ONR-VB N00014-23-1-2870 (Q.S.). The
majority of the computational calculations have been performed on the Shared University Grid
at Rice funded by NSF under Grant EIA-0216467, a partnership between Rice University, Sun
Microsystems, and Sigma Solutions, Inc., the Big-Data Private-Cloud Research Cyberinfrastructure
MRI-award funded by NSF under Grant No. CNS-1338099, and the Extreme Science and
Engineering Discovery Environment (XSEDE) by NSF under Grant No. DMR170109. H.H. acknowledges
the support of the European Research Council (ERC) under the European
Union's
Horizon 2020 research and innovation program (Grant Agreement No.\ 101020833). Work in
Vienna was supported by the Austrian Science Fund (projects I 5868-N - FOR 5249 - QUAST
and and SFB F 86, Q-M$\&$S)
and the ERC (Advanced Grant CorMeTop, No.\ 101055088). J.C. acknowledges the support of
the National Science Foundation under Grant No. DMR-1942447, support from the Alfred P.
Sloan Foundation through a Sloan Research Fellowship and the support of the Flatiron Institute,
a division of the Simons Foundation.  Six of us (L.C.,\,F.X.,\,S.S.,\,S.P.,\,J.C.,\,Q.S.)
acknowledge the hospitality of the Kavli Institute for Theoretical Physics, 
supported in part by the National Science Foundation under Grant No. NSF PHY1748958, 
during the program ``A Quantum Universe in a Crystal: Symmetry and Topology across the Correlation Spectrum".
S.S., J.C., S.P. and Q.S. acknowledge the hospitality of the Aspen Center for Physics, 
which is supported by NSF grant No. PHY-2210452.

\vspace{0.2cm}
\noindent{\bf Author contributions:}~~
Q.S. conceived the research. L.C., F.X., S.S., H.H., J.C. and Q.S. carried out 
model studies.
S.P., J.C. and Q.S. provided insights into the flat band and Kondo systems.
L.C. and Q.S. wrote the manuscript, with inputs from all authors.

\vspace{0.2cm}
\noindent{\bf Competing 
 interests:}~~
The authors declare no competing 
 interests.
 
 \vspace{0.2cm}
 \noindent{\bf Additional information:} ~~
 Correspondence and requests for materials should be addressed to 
Q.S. (qmsi@rice.edu).

\newpage

\begin{figure}[h]
\centering
\includegraphics[width=0.92\columnwidth]{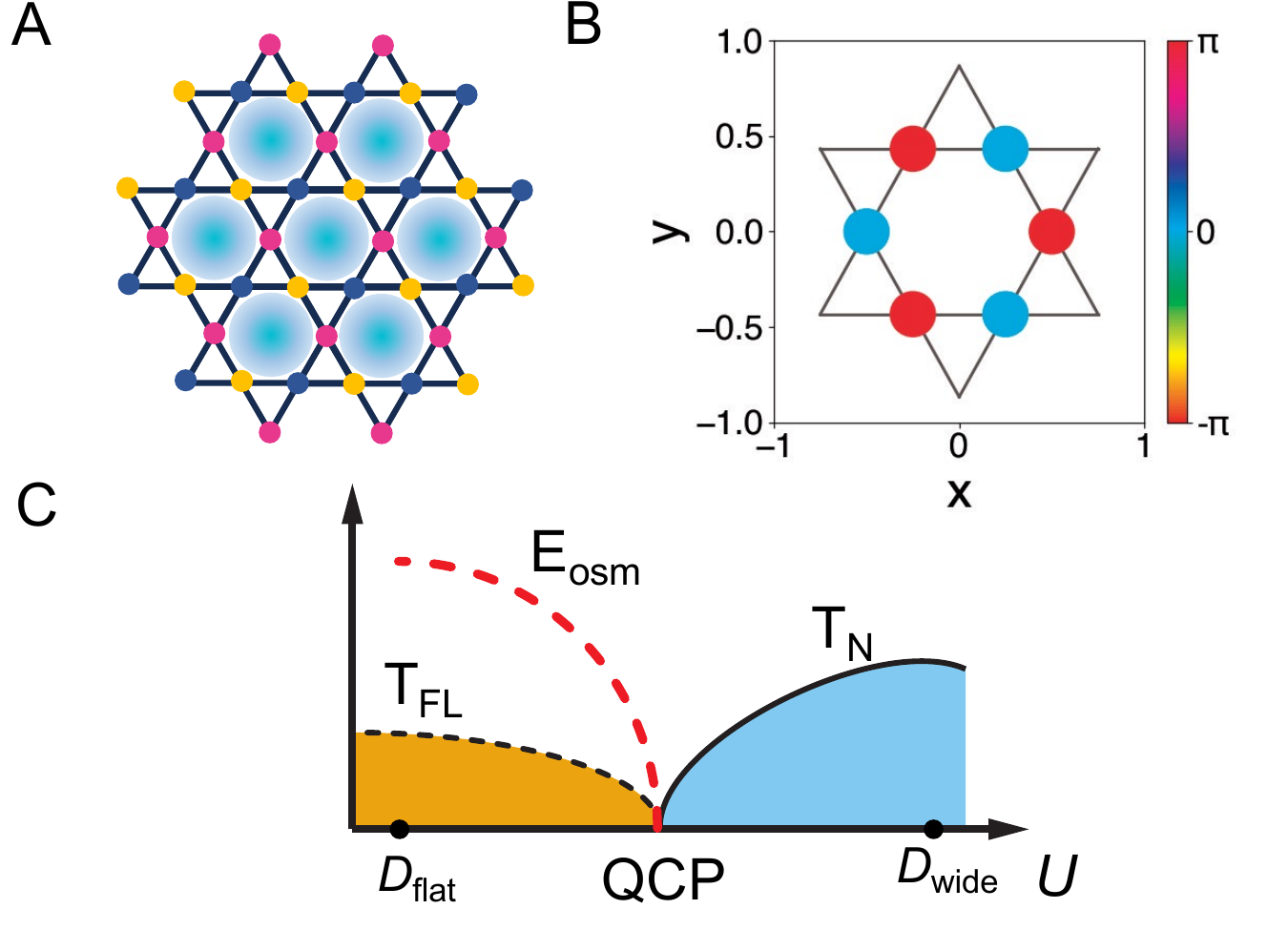}
\caption{ \textbf{Illustration of the lattice geometry, compact localized 
state, Wannier centers, and the qualitative phase diagram.}
{\bf A}, Geometry of the kagome lattice with three sites per unit cell.  The Wannier centers of the emergent Wannier orbitals are marked by the light-blue circles (centered at Wyckoff position 1a).
{\bf B}, The real space representation of the {compact localized state} for the flat band in the single orbital kagome system. 
{\bf C}, The qualitative phase diagram featuring a selective Mott transition of the molecular orbitals, with an associated energy scale for ``orbital-selective Mott" (osm) crossover (that turns into a phase transition at zero temperature).
}
\label{fig:illu}
\end{figure}

\newpage
\begin{figure}[h]
\centering
\includegraphics[width=0.95\columnwidth]{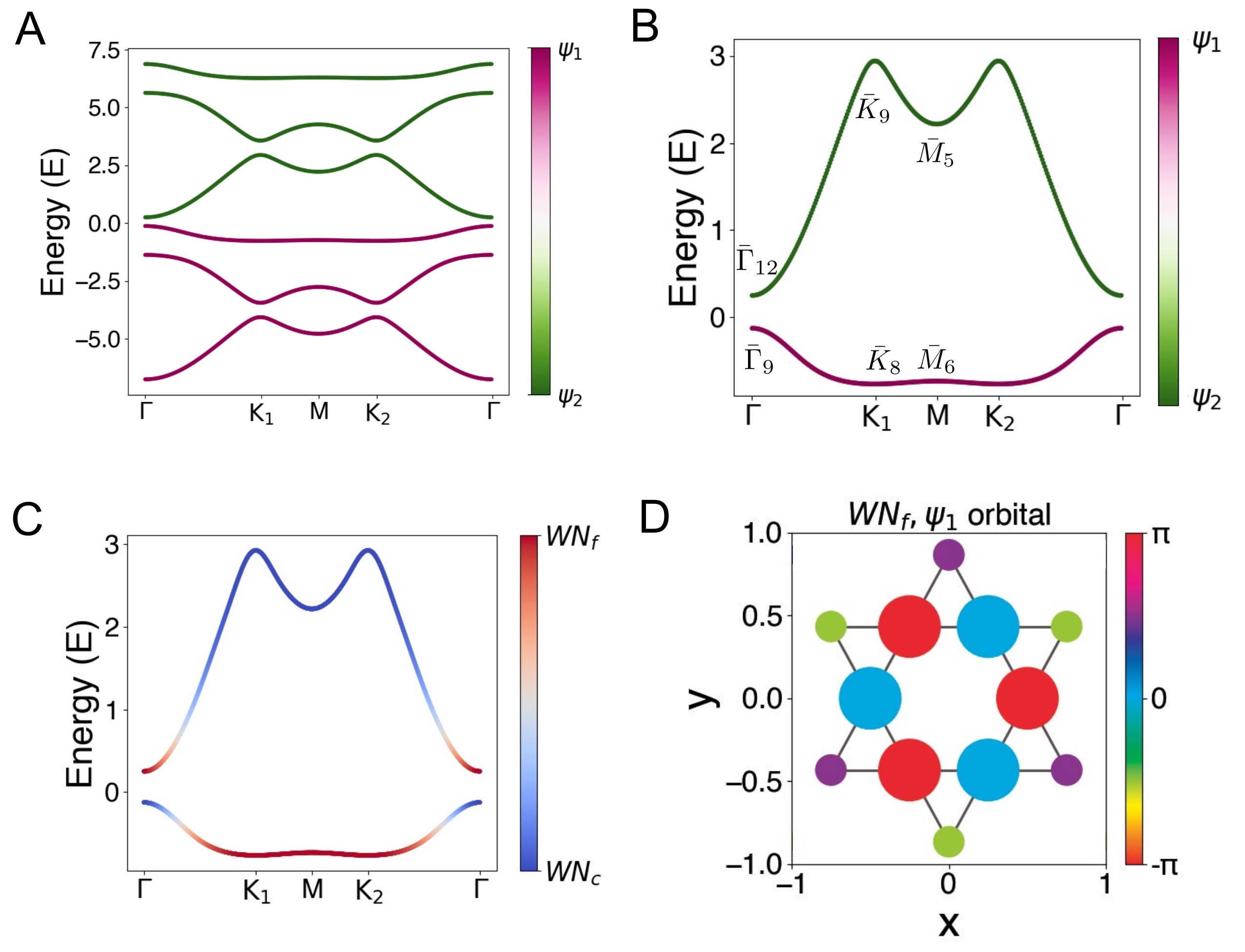}
\caption{ \textbf{Noninteracting bandstructure.} {\bf A}, The band structure of the two-orbital model in the original kagome lattice, with the parameter setting described in the Methods. 
{\bf B}, The band structure for the middle two bands in the original Hamiltonian. The little group representations at high symmetry points are marked. 
{\bf C}, The dispersion of the effective Hamiltonian using two Wannier orbitals, with the color indicating the ratio of the Wannier orbital components. {\bf D}, The shape of the Wannier orbital for $WN_{f}$ ($\psi_1$ component) in real space. The 
size (color) of the dot indicates the 
square root of the density (phase) of the Wannier function. (The density of the Wannier function is defined as  $|w(r)|^2$.)
}
\label{fig:bd}
\end{figure}

\newpage

\newpage

\newpage


\setcounter{figure}{0}
\setcounter{equation}{0}
\makeatletter
\renewcommand{\thefigure}{S\@arabic\c@figure}
\renewcommand{\theequation}{S\arabic{equation}}

\noindent{\LARGE\bf{Supplementary Materials}}
\\

\noindent{\bf\Large
Metallic quantum criticality 
enabled by flat bands in a kagome lattice
}

\vskip 0.3 cm

\noindent 
Lei\ Chen,  Fang\ Xie,  Shouvik\ Sur,  Haoyu Hu, 
Silke\  Paschen, Jennifer Cano, Qimiao Si

\vskip 0.3 cm


\noindent
\noindent Figs.\,S1 to S7
\\
Tables \,S1 to S5
\\
References (25,35,36, 42,58,59, see the above)


\section*{}

\subsection*{A. Symmetry analysis and band structure in the original Hamiltonian}\label{Sec:symm}

\begin{figure}[h]
\centering
\includegraphics[width=0.8\columnwidth]{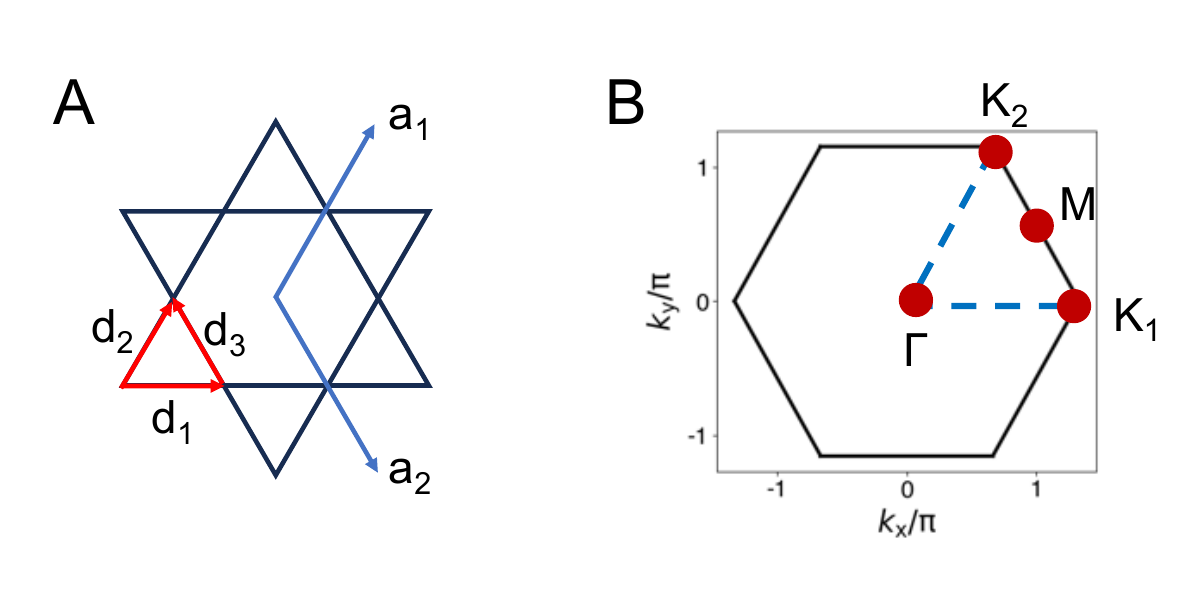}
\caption{ \textbf{Lattice structure.}
{\bf A,} 
The kagome lattice, with 
the basis vectors $\bm{a}_1=\frac{1}{2}\hat{x} + \frac{\sqrt{3}}{2}\hat{y} $ and $\bm{a}_2=\frac{1}{2}\hat{x} -\frac{\sqrt{3}}{2}\hat{y}$. Here $\hat{x}$ and $\hat{y}$ represent the unit vectors in 
Cartesian coordinate. The displacements between the sublattices are denoted by $\bm{d}_{i}$. {\bf B,} The first Brillouin zone of the kagome lattice, with some high symmetry points in the Brillouin zone marked.
}
\label{fig:org}
\end{figure}

The non-interacting Hamiltonian governing the model on the original kagome lattice is described by 
Eqs.\,(3-5) (see the Methods). This model exhibits crystalline symmetry characterized by the space group $P6/mmm$ (SG191), which preserves both the time reversal and inversion symmetries. Consequently, the bands  possess a twofold degeneracy throughout the Brillouin zone (BZ). Moreover, the $C_{2z}$ symmetry on the kagome lattice prevents any in-plane SOC, thereby maintaining the $U(1)$ spin rotational symmetry. As a result, the spin-up and spin-down sectors can be treated independently. At each kagome sublattice, which has a $D_{2h}$ site symmetry group, we include two orbitals that transform as $\bar{E}_{g}$ and $\bar{E}_{u}$ irreps. They are denoted as $\psi_{1}$ and $\psi_{2}$ respectively. 

We next turn to 
wave vector space and discuss the little-group representations at high symmetry points. To analyze the EBRs, we utilize the notation provided by the BANDREP application of the Bilbao Crystallographic Server (BCS)~\cite{BCS}.
The irreps in $\bar{E}_{g}\up G$ and $\bar{E}_{u}\up G$ from Wyckoff position $3f$ are ($\bar{\Gamma}_7$, $\bar{\Gamma}_8$, $\bar{\Gamma}_9$, $\bar{K}_7$, $\bar{K}_8$, $\bar{K}_9$, $\bar{M}_5$, 2$\bar{M}_6$) and ($\bar{\Gamma}_{10}$, $\bar{\Gamma}_{11}$, $\bar{\Gamma}_{12}$, $\bar{K}_7$, $\bar{K}_8$, $\bar{K}_9$, 2$\bar{M}_5$, $\bar{M}_6$), respectively. The specific energy order of these irreps are dependent on the parameters of the tight-binding model. However, as we gradually reduce the strength of the SOC, the band structure should converge to that of the simplest spinless kagome model with a nearest-neighbor hopping, which has band touching points corresponding to two dimentional irreps. Therefore, when the SOC is smaller than the nearest-neighbor hopping strength, we observe certain relative energy relations in the EBRs. Specifically, for $\bar{E}_g\up G$, $\bar{\Gamma}_9$ is closer to $\bar{\Gamma}_7$, and $\bar{K}_7$ is closer to $\bar{K}_9$. Similarly, for $\bar{E}_u\up G$, $\bar{\Gamma}_{11}$ is closer to $\bar{\Gamma}_{10}$ and $\bar{K}_7$ is closer to $\bar{K}_9$.

We have determined the irreps at each high symmetry point.  The corresponding results are summarized in Table \ref{tab:sym_doub}. Each row of the table provides information about the irreps at high symmetry points of the isolated double degenerate bands, ordered from the top to the bottom according to the energy hierarchy as depicted in 
Fig.\,\ref{fig:bd}A. 

\begin{table}[h]
\centering
\begin{tabular}{c | c | c | c | c }
\hline 
Wyckoff & EBRs & $\Gamma$ irreps & $K$ irreps & $M$ irreps \\
\hline \hline \xrowht{8pt}
\multirow{6}{*}{$3f$} & \multirow{3}{*}{$\bar{E}_u \up G$}  & $\bar{\Gamma}_{11}$ & $\bar{K}_{8} $ & $\bar{M}_5$\\
                   &  &  $\bar{\Gamma}_{10}$ & $\bar{K}_{7} $ & $\bar{M}_6$        \\ 
                  &   & $\bar{\Gamma}_{12}$ & $\bar{K}_{9} $ & $\bar{M}_5$          \\ \cline{2-5}\xrowht{9pt}
                  &  \multirow{3}{*}{$\bar{E}_g \up G$} & $\bar{\Gamma}_9$ & $\bar{K}_8$ & $\bar{M}_6$ \\  
                  & & $\bar{\Gamma}_7$ & $\bar{K}_7$ & $\bar{M}_5$\\
                  & & $\bar{\Gamma}_8$ & $\bar{K}_9$ & $\bar{M}_6$  \\
\hline
\end{tabular}
\caption{Band representations and the corresponding irreps at high symmetry points for the spinful Hamiltonian that follows $\mathrm{SG191}$ with orbitals at Wyckoff position $3f$.  Each row corresponds to the irreps of an isolated energy band, and the ordering is associated with the bands in Fig.\,\ref{fig:bd}A.}
\label{tab:sym_doub}
\end{table}

\begin{table}[h]
\centering
\begin{tabular}{c | c | c | c | c }
\hline 
Wyckoff & EBRs & $\Gamma$ irreps & $K$ irreps & $M$ irreps \\
\hline \hline\xrowht{8pt}
\multirow{2}{*}{1a} & $\bar{E}_{1u} \up G$ & $\bar{\Gamma}_{12}$ & $\bar{K}_8$ & $\bar{M}_6$  \\
\cline{2-5}\xrowht{8pt}
                    & $\Bar{E}_{1g} \up G$ & $\bar{\Gamma}_9$ & $\bar{K}_9$ & $\bar{M}_5$  \\

\hline
\end{tabular}
\caption{EBRs and the connection relation for $\mathrm{SG191}$ at Wyckoff $1a$.}
\label{tab:sym_doub_1a}
\end{table}

\begin{figure}[h]
\centering
\includegraphics[width=0.85\columnwidth]{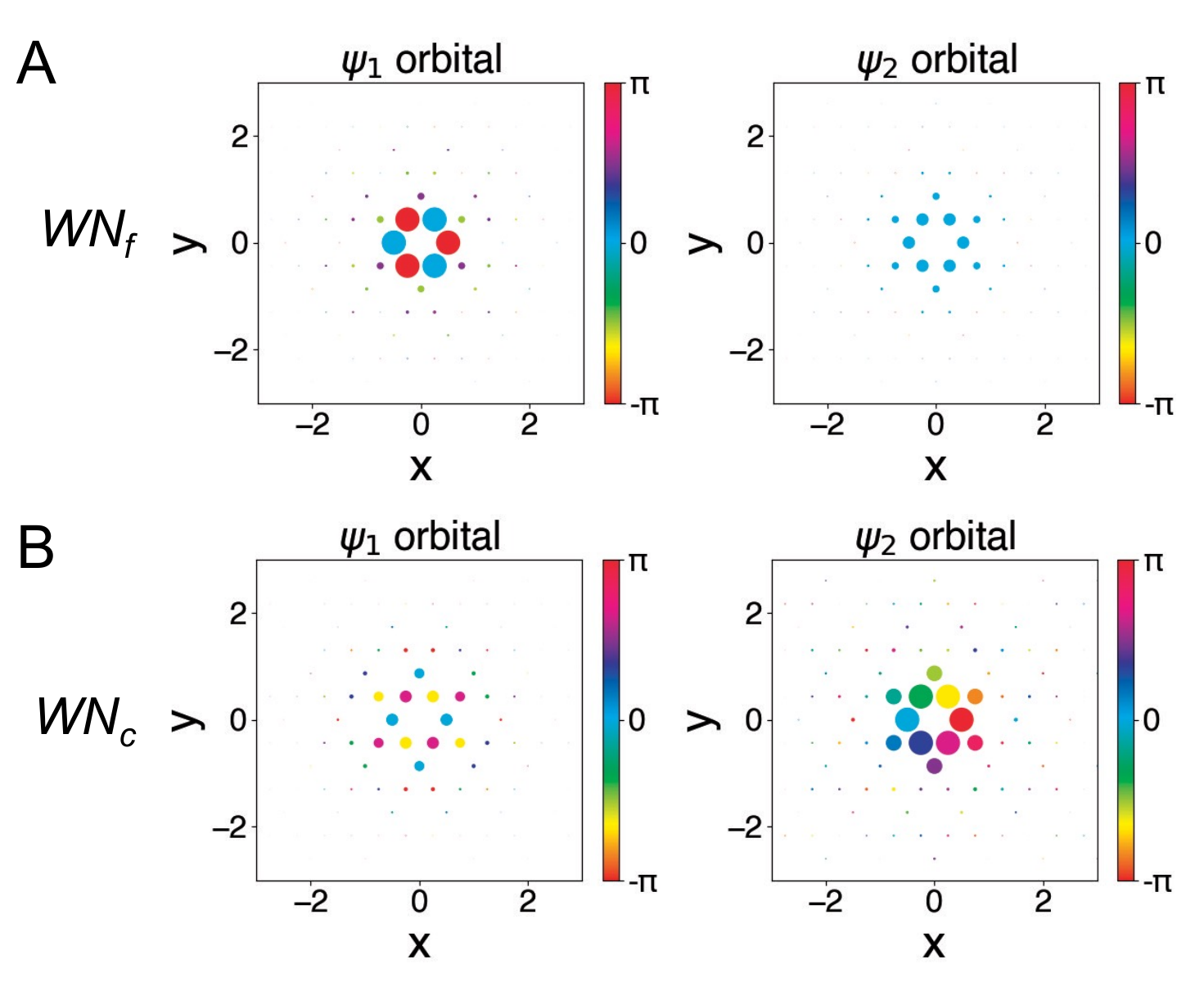}
\caption{ \textbf{Emergent Wannier orbitals.} The shape of the Wannier orbital of the spin-$\up$ sector for {\bf A, } $WN_f$ and {\bf B, } $WN_c$ in real space. The size (color) of the dot represents the density (phase) of the Wannier function in the $\psi_1$ and $\psi_2$ components.
}
\label{fig:wn}
\end{figure}

\begin{figure}[h]
\centering
\includegraphics[width=0.7\columnwidth]{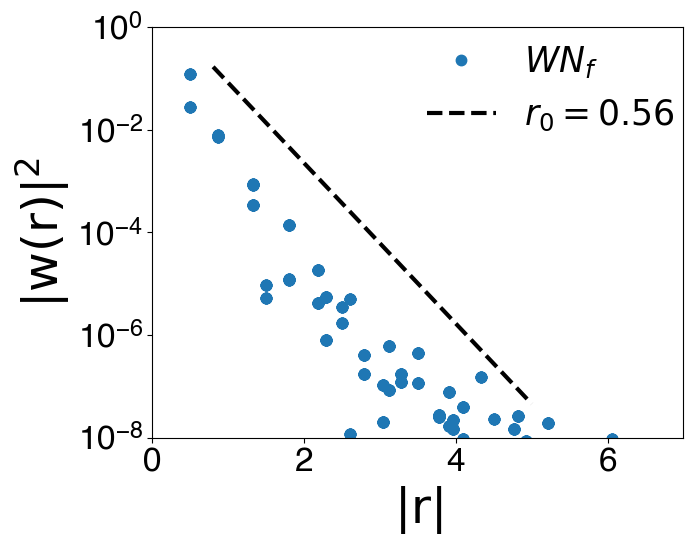}
\caption{ \textbf{ Decaying of the Wannier function density.}  The Wannier functions of the $f$ electron decay exponentially versus distance.  
}
\label{fig:wn_decay}
\end{figure}

\subsection*{B. Wannier construction}

We focus on the middle two bands and construct the Wannier orbitals.  Individually, each of
the middle two bands 
has nonzero spin Chern number and, thus,
suffers from topological obstruction~\cite{Soluyanov2011}. However, this obstruction can be circumvented if we Wannierize them together. This can be understood, if we switch over the irreps at the $\Gamma$ point for the middle two bands. As shown in Table~\ref{tab:sym_doub_1a}, after the exchange, the new sets of the little group irreps follow the EBRs $\bar{E}_{1g}\up G$ and $\bar{E}_{1u} \up G$ at Wyckoff position $1a$, respectively. In other words,  we can construct two Wannier orbitals, whose symmetry transformation follows $\bar{E}_{1g}$ and $\bar{E}_{1u}$ irreps, to represent the middle two bands. A distinct feature of these Wannier orbitals is that they are molecular orbitals. Since the new Wannier centers are at Wyckoff position 1a, they form a triangular lattice. 

We now describe the construction of the Wannier orbitals. Because the system has $U(1)$ spin rotational symmetry, it is easier to focus on a single spin species to optimize the wave function. The Wannier function for the other spin component  can be directly obtained by a Hermitian conjugate. In the following, we focus on the spin-$\up$ sector.
We first introduce two trial wavefunctions and use the projection method to determine the Wannier orbitals with the middle two bands. Based on the obtained Wannier functions, we use 
the Wannier90 software~\cite{Pizzi2020} to further optimize them by minimizing the localization functional as defined in Ref.~\cite{Vanderbilt2012-RMP}. 
The initial trial wavefunctions follow the symmetries as deduced above. In other words, for the trial wavefunctions
\begin{equation}
\begin{aligned}
        | \mathrm{Trial}; f, \Rbf \rangle &= \sum_{\bm{r}, i} e^{-|\rbf-\Rbf|/r_0} w_i^f \psi_{\rbf,i}^{\dagger} | 0 \rangle, \\
        | \mathrm{Trial}; c, \Rbf \rangle &= \sum_{\bm{r}, i} e^{-|\rbf-\Rbf|/r_0} w_i^c \psi_{\rbf,i}^{\dagger} | 0 \rangle \\
\end{aligned}
\end{equation}
the phase factors $w_i^f$ and $w_i^c$ follow the symmetry properties of the single spin sector. 
The obtained Wannier orbitals are shown in Fig.\,\ref{fig:wn}A,B, in which the components in the original $\psi_1$ and $\psi_2$ orbitals are separated into two panels. The size (color) of the dot represents the density (phase) of the Wannier function.
{The system preserves the $C_6$ rotational symmetry. Consequently, the Wannier functions can be characterized by angular momentum determined from $C_6$ eigenvalues. According to the phases shown in Fig.~\ref{fig:wn}, $WN_f$ has the angular momentum $L=3$, which is the same as the CLS as shown in Fig.~\ref{fig:illu}B. 
At the same time, $WN_c$ has the angular momentum $L=4$.}  
As shown in Fig.\,\ref{fig:wn_decay}, the obtained Wannier function is exponentially localized. {By fitting the envelope of the Wannier function density with $|w(r)|^2\sim Ae^{-2r/r_0}$, we obtain the decaying length scale for the $WN_f$ to be $r_0
=0.56$.}
The corresponding decaying length scale for the more extended Wannier function, $WN_c$, is somewhat larger, $r_0=0.9$.
The band structure using the new tight-binding parameters in the Wannier basis is plotted in 
Fig.\,\ref{fig:bd}C. One can easily observe that most parts of the flat band are dominated by the Wannier orbital $WN_f$.  

\begin{figure}[t!]
\centering
\includegraphics[width=0.8\columnwidth]{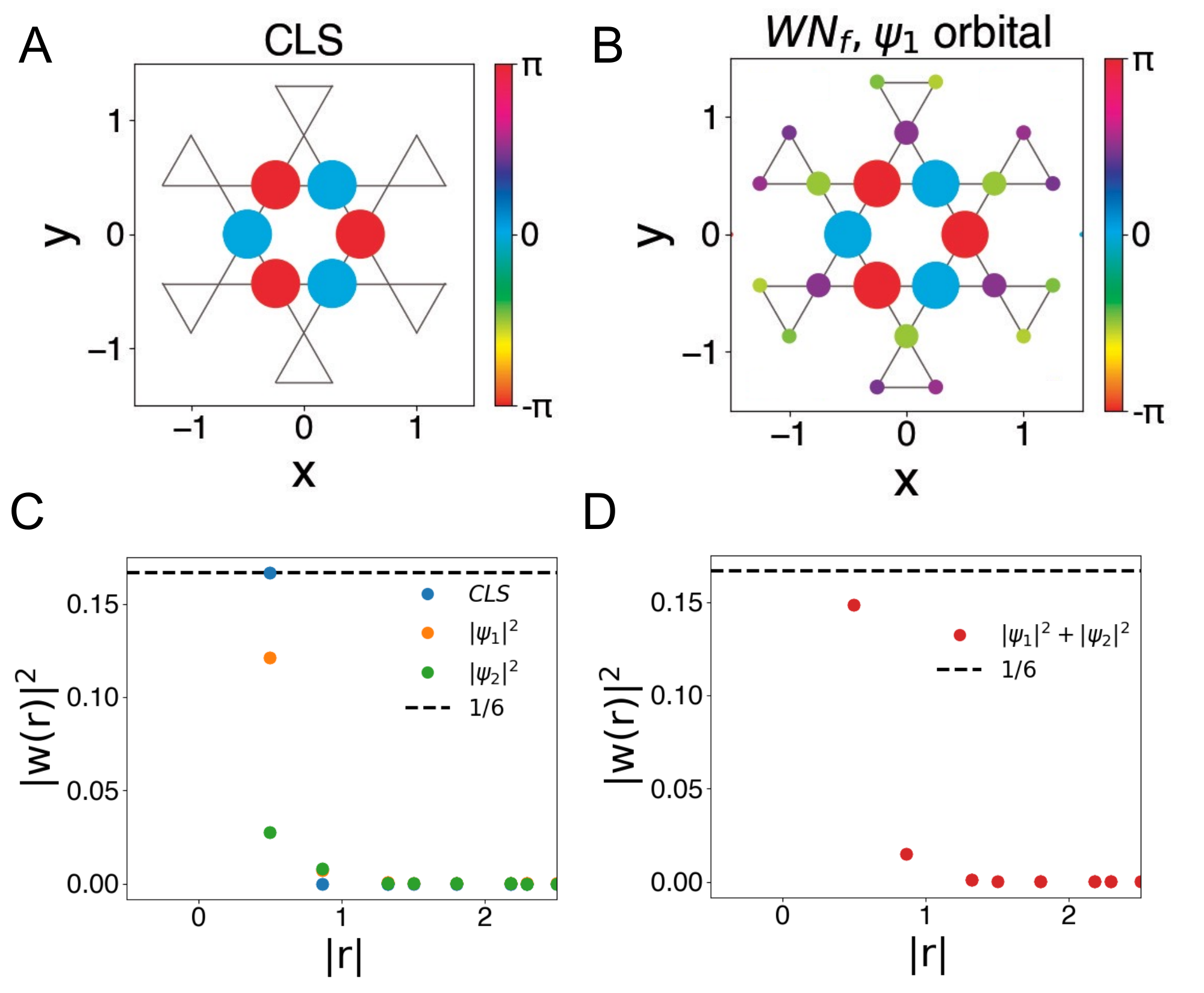}
\caption{ { \textbf{Comparison between $WN_f$
and the compact localized state.} Real space representation of {\bf A, } the compact localized state (CLS) and {\bf B, } the tight Wannier function $WN_f$
projected on to the 
$\psi_1$ orbital. Here, 
size (color) of the dot represents the square root of the density (phase). They share the same phases for the six sites on the middle hexagon, which means they 
transform the same way under symmetry. {\bf C, } Decaying of the Wannier function density for the 
compact localized state 
and the $WN_f$ in different components, and {\bf D, } the total Wannier function density of $WN_f$ versus distance. The density of the 
compact localized state is confined to the middle hexagon.
The density of $WN_f$ 
has an exponentially decaying tail.
}
}
\label{fig:cls}
\end{figure}

\subsection*{C. Relation between $WN_f$ and the compact localized state}

The compact localized state,
illustrated in Fig.\,\ref{fig:illu}B and Fig.\,\ref{fig:cls}A, can be derived from the non-interacting Hamiltonian with nearest-neighbor
hoppings on a kagome lattice without spin-orbit coupling~\cite{Bergman2008}. The 
compact localized state centered on a particular hexagon exhibits nonzero amplitudes on only the six sites of the hexagon [See Fig.\,\ref{fig:cls}A,C]. 
While the 
compact localized state are exact eigenstates making up the flat band, they are not orthogonal, and, more importantly, they do not constitute a complete basis for the flat band since they are not linearly independent \cite{Bergman2008}.
In contrast, the Wannier orbitals, $WN_f$, despite sharing the same point group symmetry and center as the 
compact localized state (see Fig.\,\ref{fig:cls}A,B),
are both orthogonal (see below) and linearly independent, thus forming a complete basis.
Furthermore, as depicted in Fig.\,\ref{fig:cls}D, the Wannier function density ($|w(r)|^2$) of the $WN_f$ are predominantly localized within a single hexagon as well. 
Although the Wannier orbitals mostly comprise the states from the flat band, they also have weight on the dispersive band near $\Gamma$
(see Fig.\,\ref{fig:bd}C), as required by the topological obstruction.
The orthogonality of the $WN_{f}$ is assured by the Wannierization procedure. 
This is to be contrasted with the compact localized states, which are not orthogonal: For example, the overlap between two nearest neighbor compact localized states
is $-\frac{1}{6}$.

\begin{figure}[t!]
\centering
\includegraphics[width=1\columnwidth]{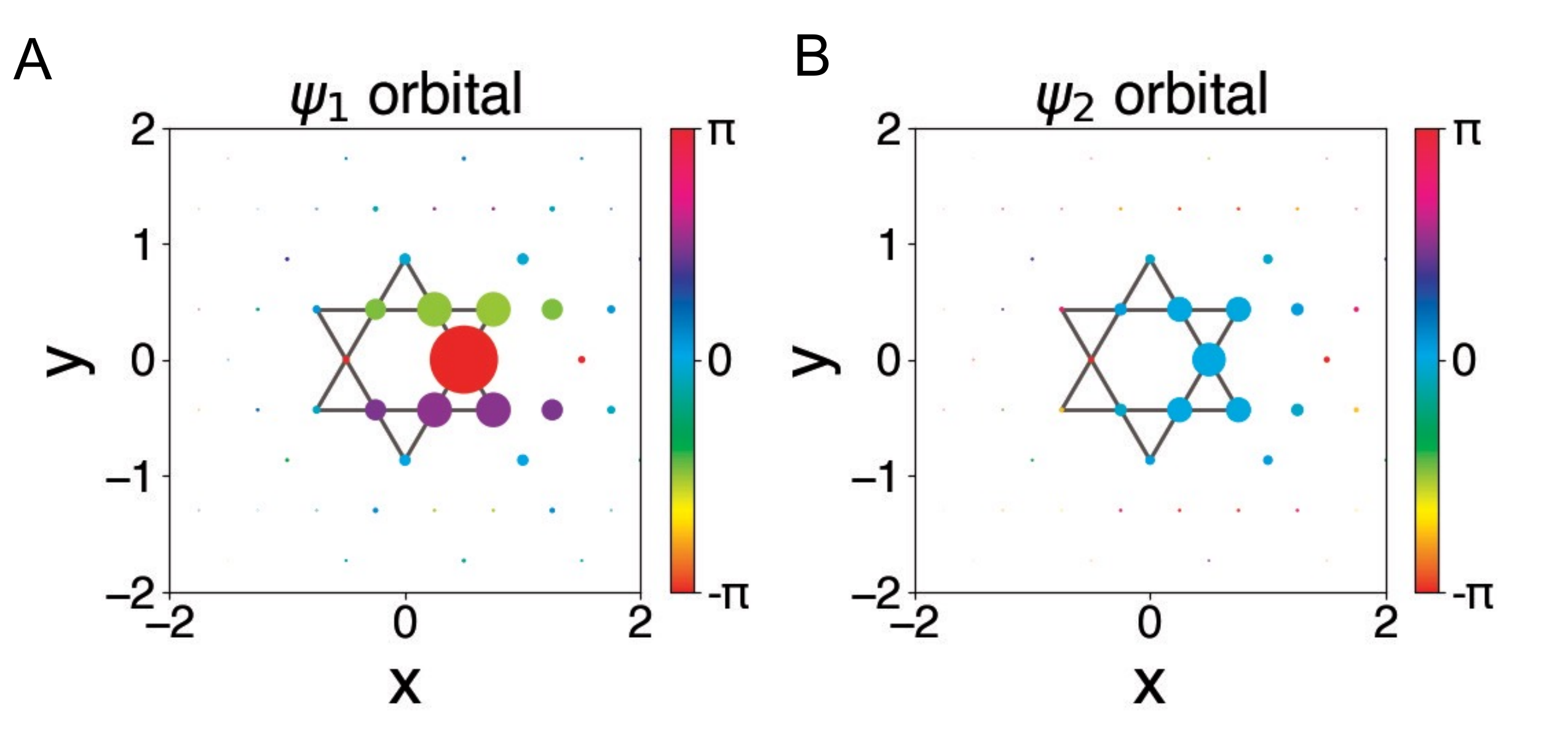}
\caption{ Real space representation of the overlapping functional $\mathcal{O}^{\alpha\alpha}(x, y)$ in {\bf A, } orbital $\psi_1$ and {\bf B, } orbital $\psi_2$, for two nearest neighbor $WN_f$'s  \jc{at the unit cell ${\bf R}_0 = {\bf 0}$ and ${\bf R}_1 = \abf_1 + \abf_2$ as indicated above in Eq. ~\ref{eq:overlap} }.
Here the axis $x$ and $y$ label the positions of the kagome sites in the Cartesian coordinate. }
\label{fig:wRR}
\end{figure}

To explicitly illustrate this point, we evaluate the overlap between two nearest neighbor $WN_{f}$ states. 
We consider the unit cells 
centered at an origin,
$\mathbf{R}_0={\bf 0}$ and at 
$\mathbf{R}_1=\abf_1+\abf_2$;
in other words, one anchored by
a hexagon and the other centered at the hexagon to its immediate right.
The overlap between the $WN_{f}$ states
associated with these unit cells
is specified by
\begin{equation}\label{eq:overlap}
\begin{aligned}
\mathcal{O} &= \sum_{
{\rm sites},\alpha\beta} \mathcal{O}^{\alpha\beta}(x, y) \, ,\\
\mathcal{O}^{\alpha\beta}(x,y) &= \left[ WN_f(\mathbf{R}_0) \right]^*_{\alpha,(x,y)}\left[ WN_f(\mathbf{R}_1) \right]_{\beta,(x,y)} \, ,
\end{aligned}
\end{equation}
where $WN_f(\mathbf{R})$ indicates the Wannier function $WN_f$ centered at $\mathbf{R}$ and the subscripts 
``$\alpha, (x,y)$" indicate its components on the orbital $\psi_\alpha$ at the kagome site $(x,y)$ in Cartesian coordinates.
Note that $\mathcal{O}^{\alpha\beta}$ is only non-zero when $\alpha=\beta$, 
due to the orthogonality of the two orbitals $\psi_1$ and $\psi_2$.
The specific values of the overlap functional, $\mathcal{O}^{\alpha\alpha}(x, y)$, are displayed in Table\,\ref{tab:wRR}, and the corresponding real space representation is depicted in Fig.\,\ref{fig:wRR}. 
Table\,\ref{tab:wRR} also 
shows $\mathcal{O}^{\alpha\alpha}
=\sum_{\rm sites} \mathcal{O}^{\alpha\alpha} (x,y)$.
The sum 
$\mathcal{O}^{11}\approx -0.1$;
the magnitude 
is of the same order as 
the 
overlap between two nearest neighbor 
compact localized states
($-1/6$). However, in $WN_f$, $\mathcal{O}^{22}$ is equal 
to $ -\mathcal{O}^{11}$ (see Table\,\ref{tab:wRR}).
The contributions from the two orbitals 
exactly 
cancel out, 
leading to the overall overlap $\mathcal{O}=0$, i.e.
the orthogonality between the $WN_f$ states.

These observations emphasize
the connection between the Wannier orbital $WN_f$ and the compact localized state. More importantly, they highlight the
distinction between the two states.

\begin{table}[h]
    \centering
    \begin{tabular}{c|c|c|c|c}
    \hline\xrowht{10pt}
         $(x, y)$ &  $(\frac{1}{2}, 0)$ & $(\frac{1}{4}, \frac{\sqrt{3}}{4} )$  & $(\frac{1}{4}, -\frac{\sqrt{3}}{4} )$ & $(\frac{3}{4}, \frac{\sqrt{3}}{4} )$ \\
         \hline\xrowht{8pt}
         $\mathcal{O}^{11}(x, y)$ & $-0.1211$  & $0.0293e^{-i\pi/2}$ & $0.0293e^{i\pi/2}$ & $0.0293e^{-i\pi/2}$  \\
         \hline\xrowht{8pt}
         $\mathcal{O}^{22}(x, y)$ & $0.02757$ & $0.014734$ & $0.014734$ & $0.014734$ \\
         \hline\xrowht{10pt}
         $(x, y)$ & $(\frac{3}{4}, -\frac{\sqrt{3}}{4} )$ & $(-\frac{1}{4}, \frac{\sqrt{3}}{4})$ & $(\frac{5}{4}, -\frac{\sqrt{3}}{4})$ & $(\frac{1}{4}, \frac{\sqrt{3}}{4})$  \\
         \hline\xrowht{8pt}
         $\mathcal{O}^{11}(x, y)$ & $0.0293e^{i\pi/2}$ & $0.01e^{-i0.471\pi}$  & $0.01e^{i0.471\pi}$ & $0.01e^{-i0.471\pi}$ \\
         \hline\xrowht{8pt}
         $\mathcal{O}^{22}(x, y)$ & $0.014734$ & $0.00307e^{i0.0477\pi}$ & $0.00307e^{-i0.0477\pi}$ & $0.00307e^{i0.0477\pi}$  \\
         \hline\xrowht{10pt}
         $(x, y)$ & $(\frac{1}{4}, -\frac{\sqrt{3}}{4})$ & $(1, \frac{\sqrt{3}}{2})$ &   $(0, -\frac{\sqrt{3}}{2})$ &  $(0, \frac{\sqrt{3}}{2})$   \\
         \hline\xrowht{8pt}
         $\mathcal{O}^{11}(x, y)$ & $0.01e^{i0.471\pi}$ & $0.0025e^{-i0.003\pi}$  & $0.0025e^{i0.003\pi}$ & $0.0025e^{-i0.003\pi}$  \\
         \hline\xrowht{8pt}
         $\mathcal{O}^{22}(x, y)$ & $0.00307e^{-i0.0477\pi}$ & $0.00164e^{-i0.047\pi}$ & $0.00164e^{i0.047\pi}$ & $0.00164e^{-i0.047\pi}$ \\
         \hline\xrowht{10pt}
         $(x, y)$ & $(1, -\frac{\sqrt{3}}{2})$ & $(-\frac{1}{2}, 0)$ & $(\frac{3}{2}, 0)$\\
         \hline\xrowht{8pt}
         $\mathcal{O}^{11}(x, y)$ & $0.0025e^{i0.003\pi}$ & $-0.0008$ & $-0.0008$ \\
         \hline\xrowht{8pt}
         $\mathcal{O}^{22}(x, y)$ & $0.00164e^{i0.047\pi}$ & $-0.0005$ & $-0.0005$\\
         \hline\hline\xrowht{12pt}
         $\sum_{\rm sites}\mathcal{O}^{11}$ & $\sum_{\rm sites}\mathcal{O}^{22}$ & $\sum_{\rm sites} \left ( \mathcal{O}^{11} + \mathcal{O}^{22} \right )$ \\
         \hline\xrowht{8pt}
         $-0.10287$ & $0.10287$ & $0$\\
         \hline
    \end{tabular}
    \caption{ Parts of the overlapping functional $\mathcal{O}^{\alpha\alpha}$ for two nearest neighbor $WN_{f}$'s 
    on the kagome sites 
    $(x, y)$. The sum over all the kagome sites equal to $0$, which means the Wannier orbitals are orthogonal to each other. 
    A graphical representation of the same distribution is provided in Fig.~\ref{fig:wRR}. 
    }
    \label{tab:wRR}
\end{table}

\subsection*{D. Effective Andersion lattice model}
Upon performing the Wannier construction, we can project the original Hamiltonian, defined on the kagome lattice, onto the newly constructed basis to obtain an effective extended Anderson model. The effective Hamiltonian retains all the symmetries from $\mathrm{SG191}$. The specific form of the kinetic part is described by Eq.\,(6) in the Method section with the parameters listed in Table~\ref{tab:hop}. 
The hybridization between the Wannier functions $WN_c$ and $WN_f$ is offsite,
since these two Wannier functions belong to different irreps.

We further project the local Hubbard interaction onto the newly developed Wannier basis, which decays exponentially. It takes the form:
\begin{equation}
\begin{aligned}
    H_{int} &= H_{int}^{0} +  H_{int}^{1} + ... \\
    & = \sum_{i} \Big[ u^{f}_0 n_{f\up}^in_{f\dn}^i + u^c_{0} n^i_{c\up}n^i_{c\dn} + F_0\left(n^i_{f\up}n^i_{c\dn} + n^i_{f\dn}n^i_{c\up} \right) \\
    & + \left(F_0-J_H^{0}\right) \left(n_{f\up}^in_{c\up}^i + n_{f\dn}^in_{c\dn}^i \right) + J_H^{0} \left( f_{i\up}^{\dagger}f_{i\dn}^{\dagger} c_{i\dn}c_{i\up} + h.c. \right) \Big] \\
    & + \sum_{\langle ij \rangle} \Big[ u^{f}_1 n_{f}^i n_f^j - 4J_{H_1}^{f} \bm{S}^{i}_f\bm{S}^{j}_f + u^{c}_1 n_{c}^i n_c^j - 4J_{H_1}^{c} \bm{S}^{i}_c\bm{S}^{j}_c \\
    &+ u^{fc}_1 n_{f}^i n_c^j - 4J_{H_1}^{fc} \bm{S}^{i}_f\bm{S}^{j}_c + ... \Big] +... \, ,
\end{aligned}
\end{equation}
where $u_0^f$ ($u_0^c$) is the on-site Hubbard interaction of the $f$ ($c$) electrons. In addition, $F_0$ and $J_H^0$ are the on-site interorbital density-density and Hund's interactions. 
We focus on the case with the $WN_f$ orbitals near half-filling, for which the $f$-$c$ density-density interaction $F_0$ is unimportant~\cite{Hu_fb_2022}.
$n_{\tau}^{i} = n_{\tau\up}+n_{\tau\dn}$ and $\bm{S}^{i}_{\tau} = \psi_{\tau}^{\dagger} \frac{\bm{\sigma}}{2} \psi_{\tau}$, with $\tau=f/c$, are the density and spin for the orbital $\tau$ at site $i$, respectively. And $u_1^{\alpha}$ ($J_{H_1}^{\alpha}$), with $\alpha=f/c/fc$, represents the nearest neighbor intra/inter-orbital density-density (Hund's) interactions.
The values of the $u_{1}^{\alpha}$ is about $1/6$ of the $u_{0}^{\alpha}$
This is because the Wannier function $WN_{f}$ is mostly concentrated on the middle six sites of a hexagon. And for the nearest neighbor two $WN_{f}$, they share one corner of the hexagon, leading to a ratio of $1/6$.
Notice that the interaction Hamiltonian only has the $U(1)$ symmetry: The $
{S}^{i}_{x} 
{S}^i_x+ 
{S}^{i}_{y} 
{S}^i_y$ terms are forbidden by the symmetry of the Wannier function. The values of these parameters are listed in Table~\ref{tab:u} in unit of the original $U$. The on-site Hubbard
interaction on the $f$ orbital is the largest.
While other types of interactions can arise during the projection, their magnitudes are smaller compared to the predominant on-site
interactions. 
As a result, we  only consider the on-site
interaction of the $WN_f$ orbital in the subsequent EDMFT calculation.

\begin{table}[h]
    \centering
    \begin{tabular}{c|c|c|c|c|c|c}
    \hline
         $(i,j)$ &  $\pm(1,0)$ & $\pm(0,1)$ & $\pm(1,1)$ & $\pm(2,0)$ & $\pm(0,2)$& $\pm(2,2)$\\
         \hline
         $t^{f}_{ia_1+ja_2}$ & $0.14119$ & $0.14119$ & $0.14119$ & $-0.01714$ &  $-0.01714$  &  $-0.01714$  \\
         \hline\hline
         $(i,j)$ &  $\pm(1,0)$ & $\pm(0,1)$ & $\pm(1,1)$ & $\pm(2,0)$ & $\pm(0,2)$& $\pm(2,2)$ \\
         \hline
         $t^{c}_{ia_1+ja_2}$ & $-0.35511$ & $-0.35511$ & $-0.35511$ & $0.066$ &  $0.066$  &  $0.066$\\
         \hline\hline
         $(i,j)$ & $(1, 0)$ & $(0, -1)$ & $(-1, -1)$ & $(-1, 0)$ & $(0, 1)$ & $(1, 1)$ \\
         \hline
         $V^{\up(fc)}_{ia_1+ja_2}$ & $0.212e^{i1.03}$ & $e^{i\frac{\pi}{3}}V^{\up(fc)}_{a_1}$ & $e^{i\frac{2\pi}{3}}V^{\up(fc)}_{a_1}$ & $e^{i\pi}V^{\up(fc)}_{a_1}$ & $e^{i\frac{4\pi}{3}}V^{\up(fc)}_{a_1}$& $e^{i\frac{5\pi}{3}}V^{\up(fc)}_{a_1}$\\
         \hline
         $(i,j)$ & $(1, -1)$ & $(-2, -1)$ & $(-1, -2)$ & $(-1, 1)$ & $(1, 2)$ & $(2, 1)$ \\
         \hline
         $V^{\up(fc)}_{ia_1+ja_2}$ & $0.02e^{-i1.59}$ & $e^{i\frac{\pi}{3}}V^{\up(fc)}_{a_1-a_2}$ & $e^{i\frac{2\pi}{3}}V^{\up(fc)}_{a_1-a_2}$ & $e^{i\pi}V^{\up(fc)}_{a_1-a_2}$ & $e^{i\frac{4\pi}{3}}V^{\up(fc)}_{a_1-a_2}$& $e^{i\frac{5\pi}{3}}V^{\up(fc)}_{a_1-a_2}$\\
         \hline  
         $(i,j)$ & $(2, 0)$ & $(0, -2)$ & $(-2, -2)$ & $(-2, 0)$ & $(0, 2)$ & $(2, 2)$ \\
         \hline
         $V^{\up(fc)}_{ia_1+ja_2}$ & $0.03e^{i2.078}$ & $e^{i\frac{\pi}{3}}V^{\up(fc)}_{2a_1}$ & $e^{i\frac{2\pi}{3}}V^{\up(fc)}_{2a_1}$ & $e^{i\pi}V^{\up(fc)}_{2a_1}$ & $e^{i\frac{4\pi}{3}}V^{\up(fc)}_{2a_1}$& $e^{i\frac{5\pi}{3}}V^{\up(fc)}_{2a_1}$\\ 
         \hline\hline
         & $\epsilon_{f}$ & $\epsilon_c$ \\
         \hline
         &$-0.379$ & $1.421$ \\
         \hline
    \end{tabular}
    \caption{{\bf Parameters for the noninteracting Hamiltonian in the Wannier basis.} Shown here are the intraorbital hopping parameters for the $f$ and $c$ orbitals and the hybridization between the $f$ and $c$ orbitals for the spin up component. {The three 
    rows of the ($V^{fc}$) denote the nearest, next nearest and next-next nearest hybridization between the $f$ and $c$ electrons, respectively.}
    The parameters in the spin-down component are the Hermitian conjugate of their spin-up counterparts.}
    \label{tab:hop}
\end{table}

\begin{table}[h]
    \centering
    \begin{tabular}{c|c|c|c|c|c|c|c|c|c}
    \hline\xrowht{9pt}
         $u_0^{f}$ & $u_0^{c}$ & $F_0$ & $J_H^{0}$ & $u_1^f$ & $J_{H_1}^f$ & $u_1^c$ & $J_{H_1}^c$ & $u_1^{fc}$ & $J_{H_1}^{fc}$\\
         \hline
         $0.134$ & $0.088$ & $0.103$ & $0.063$ & $0.023$ & $0.005$ & $0.022$ & $0.006$ & $0.033$ & $0.001$\\
         \hline
    \end{tabular}
    \caption{{\bf Projected interactions in the Wannier basis.} Shown here are the projected interactions
    in unit of the 
    Hubbard interaction ($U$)
    in the original Hamiltonian. 
    }
    \label{tab:u}
\end{table}

\subsection*{E. Wannier construction of the three-band model}

\begin{figure}[h]
\centering
\includegraphics[width=0.6\columnwidth]{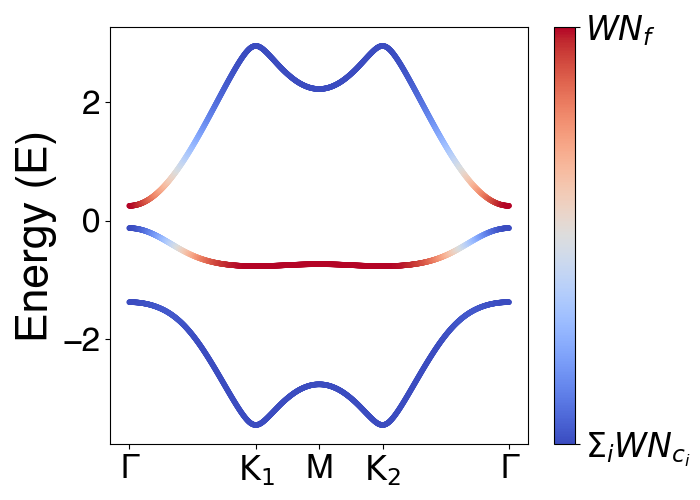}
\caption{The dispersion of the effective Hamiltonian using three Wannier orbitals, with color indicating the ratio of the orbital components. Here, the summation is over $i=1,2$, describing the two light Wannier orbitals.
}
\label{fig:bd_3orb}
\end{figure}

\begin{figure}[h]
\centering
\includegraphics[width=0.85\columnwidth]{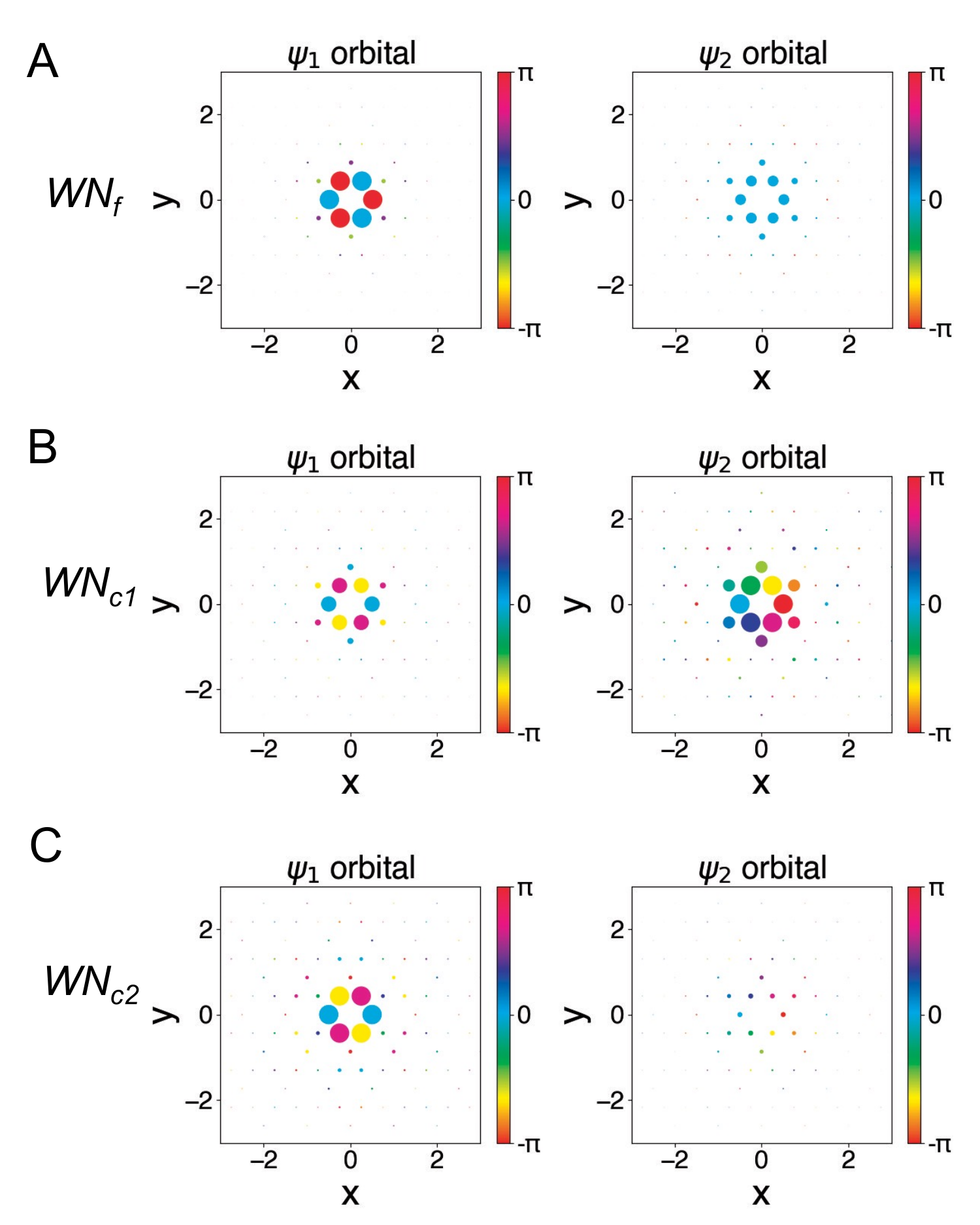}
\caption{ \textbf{Emergent Wannier orbitals of the three-band model.} The shape of the Wannier orbital of the spin-$\up$ sector for {\bf A, } $WN_f$, {\bf B, } $WN_{c_1}$ and {\bf c, } $WN_{c_2}$ in real space. The size (color) of the dot represents the density (phase) of the Wannier function in the $\psi_1$ and $\psi_2$ components. $WN_{f}$ and $WN_{c_1}$ are similar to the Wannier construction in the two orbital case, and $WN_{c_2}$ dominates the dispersive band below the flat band. 
}
\label{fig:wn3}
\end{figure}

The dispersive band below the targeted flat band is 
energetically 
not too far from the flat band. This band is topologically trivial and can be Wannierized alone. Therefore one can construct a tight binding model based on three effective Wannier orbitals with one heavy orbital ($WN_f$) and two light orbitals ($WN_{c_1}$ and $WN_{c_2}$). The obtained Wannier orbitals are depicted in Fig.\,\ref{fig:wn3}, with the fitted band structure shown in
Fig.\,\ref{fig:bd_3orb}.
In this three-orbital construction, $WN_f$ and $WN_{c_1}$ are similar to those obtained in the two-orbital model, while $WN_{c_2}$ dominates the dispersive band below the flat band.
The hybridization strength between the $f$ orbital and the two $c$ orbitals, represented by $V^{fc_1}$ and $V^{fc_2}$, depends on the specific parameters of the original Hamiltonian.
In the current parameter setting, the maximum value of $V^{fc_1}$ is approximately $0.2$. This 
is larger than the maximum value of $V^{fc_2}$, which is around $0.1$. In the low energy limit, 
a two-channel Kondo model
with unequal Kondo couplings
flows to a single-channel Kondo fixed point, which is described by the single-channel Kondo lattice model where the $f$ electrons 
hybridize with the dominantly-coupled $c_{1}$ orbital. Furthermore, the Kondo-destruction quantum criticality
is known to be insensitive to the microscopic structure of the conduction electrons~\cite{Si-Nature}. Therefore we 
focus on the two orbital model in the subsequent calculations. 

\end{document}